%% file: main.tex
\documentclass[fleqn,10pt]{wlscirep}
\usepackage[utf8]{inputenc}
\usepackage[T1]{fontenc}

\usepackage[switch]{lineno}

\usepackage{hyperref}
\usepackage{dblfloatfix}
\usepackage{float}
\usepackage{times}
\usepackage{epsfig}
\usepackage{bm}
\usepackage{breakurl}
\usepackage{graphicx, subfigure}
\usepackage{amsmath,amssymb, amsfonts, euscript, mathrsfs}
\usepackage{subeqnarray}
\usepackage{cases}
\usepackage{algorithm}
\usepackage{algorithmic}
\usepackage{fancybox}
\usepackage{upgreek}
\definecolor{lightgray}{gray}{0.9}
\definecolor{lightblue}{rgb}{0.98,0.98,1.0}

\usepackage[normalem]{ulem}

\definecolor{airforceblue}{rgb}{0.36, 0.54, 0.66}
\definecolor{arsenic}{rgb}{0.5, 0.10, 0.8}
\definecolor{darkgreen}{rgb}{0.1, 0.7, 0.1}

\newcommand{\revise}[2]{{#2}}

\newcommand{\executeiffilenewer}[3]{%
\ifnum\pdfstrcmp{\pdffilemoddate{#1}}%
{\pdffilemoddate{#2}}>0%
{\immediate\write18{#3}}\fi%
}

\graphicspath{ {./images/} }
\newcommand{\T}{\mathsf{T}}

\title{Simultaneous Topology Optimization of Differentiable and Non-Differentiable Objectives via Morphology Learning: Stiffness and Cell Growth on Scaffold}

\author[1,*]{Weiming Wang}
\author[1,*]{Yanhao Hou}
\author[1]{Renbo Su}
\author[1]{Weiguang Wang}
\author[1,$\dag$]{Charlie C.L. Wang}
\affil[1]{Department of Mechanical and Aerospace Engineering, The University of Manchester, United Kingdom\vspace{15pt}}

\affil[*]{These authors contributed equally to this work}
\affil[$\dag$]{Corresponding author (Email: changling.wang@manchester.ac.uk)}

\keywords{Structural pattern, Topology optimization, Neural network, Scaffold, Cell culture}

\input{abstract.tex}
\begin{document}

\flushbottom
\maketitle
%
%
\thispagestyle{empty}

\input{introduction}

\input{results}

\input{discussion}

\input{methods}

\section*{Data availability}
The data generated in this study will be deposited in the Github link as:\\ \url{https://github.com/W-W-M/Morphology-Learning/tree/main/data} 

\section*{Code availability}
The code used for morphology learning and the latent code based topology optimization can be accessed at Github:\\ \url{https://github.com/W-W-M/Morphology-Learning/tree/main/code} 

\bibliography{sample}




\section*{Acknowledgements}
The project is supported by the chair professorship fund at the University of Manchester and UK Engineering and Physical Sciences Research Council (EPSRC) Fellowship Grant (Ref.\#: EP/X032213/1). Weiguang Wang is also supported by Rosetrees Trust grant (Ref.\#: CF-2023-I-2$\backslash$103).

\section*{Author contributions}
Weiming Wang: Conceptualization, Methodology, Validation, Visualization, Data Collection, Software, Investigation, Formal analysis, Writing - original draft. 
Yanhao Hou: Methodology, Software, Investigation, Formal analysis, Validation, Data curation, Visualization, Writing - original draft.
Renbo Su: Investigation, Visualization, Data Curation, Software, Writing - original draft. 
Weiguang Wang: Conceptualization, Methodology, Resources, Writing - original draft, Writing - review \& editing, Supervision, Funding acquisition.
Charlie C.L. Wang: Conceptualization, Methodology, Resources, Formal analysis, Writing - original draft, Writing - review \& editing, Supervision, Funding acquisition.

\section*{Competing interests}
The authors declare no competing interests.

%




\end{document}

%% file: abstract.tex
\begin{abstract}
Topology optimization of microstructures plays a critical role in optimizing functional performance across diverse engineering applications. While metamaterials with enhanced mechanical properties -- such as hyperelasticity, energy absorption, and thermal efficiency -- are commonly designed using complex microstructural geometries and multi-physics simulations, achieving the simultaneous optimization of mechanical performance and non-differentiable objectives remains a significant challenge. In this work, we propose a novel framework for simultaneous topology optimization of differentiable and non-differentiable objectives via a data-driven morphology learning approach. The framework extracts shape patterns from a curated dataset of microstructures recognized for their superior performance in specific functional applications. To showcase the versatility of the approach, we apply it to the optimization of scaffolds for bone tissue engineering, with cell growth as a representative functional objective. By integrating learned morphology patterns into a topology optimization process, the method generates microstructures that effectively balance mechanical stiffness and biological performance, such as enhanced cell proliferation. As a case study, we demonstrate a scaffold design that improves mechanical stiffness by 29.69\% and cell growth by 37.05\% on Day 7 and 33.30\% on Day 14. This approach highlights the general applicability of the proposed framework for optimizing a broad range of engineering challenges, beyond the specific case of cell growth.
\end{abstract}

%% file: introduction.tex
\section*{Introduction}\label{sec:intro}
Geometric shape significantly influences various aspects of microstructures in scaffolds fabricated through additive manufacturing, including mechanical stiffness, heat dissipation, fluid dynamics, manufacturability, and, importantly, cell proliferation~\cite{hollister2005porous, chen2011microstructure, zhang2019three, metz2020towards}.
Traditionally, optimizing scaffold designs by adjusting the shape and topology of their microstructures has relied heavily on trial-and-error testing in both mechanical and biological experiments. Although additive manufacturing provides virtually limitless possibilities for microstructural design, this extensive design space presents both opportunities and challenges\cite{kumar2016low, wang2016topological, charbonnier2021additive}. While the potential for innovation is substantial, systematically identifying a microstructure that simultaneously optimizes both mechanical and biological performance -- such as achieving high mechanical stiffness~\cite{moroni20063d, zhang2020biomechanical, foroughi2022multi} and enhanced cell proliferation~\cite{van20223d, jin2024precision} -- remains exceptionally challenging. This difficulty is further compounded when the search must be conducted within a limited number of iterations, rendering conventional trial-and-error methods inefficient and often ineffective. 

A data-driven approach leveraging neural networks offers a powerful method for learning and encoding the shape patterns of various microstructures. By utilizing techniques such as \textit{Neural Signed Distance Functions} (named as Neural SDF) \cite{park2019deepsdf}, the complex geometries of microstructures can be captured and represented in a compact, low-dimensional latent space (e.g., 128 coefficients in our exploration) with the help of neural networks. When the values of the latent code are altered -- either systematically or randomly -- the neural network can generate new microstructures that retain similar geometric patterns to those originally learned, thus enabling the creation of designs that exhibit desirable properties\cite{hui2022neural,abbas2023deepmorpher,Guillard2024}. This approach not only reduces the computational burden of exploring vast design spaces but also facilitates the generation of novel microstructures with optimized performance characteristics, as demonstrated in recent works on neural implicit representations for 3D shapes. Leveraging natural patterns has proven advantageous in the design of engineering structures \cite{siddique2022lessons}, yet this approach has not been widely integrated into numerical optimization routines to maintain existing biological performance. Our \textbf{main insight} is that, compared to random shape exploration in unfiltered datasets, learning the latent code to parameterize the shape of microstructures within a dataset known for good biological performance significantly increases the probability of reconstructing or generating new structures with similarly good biological performance using the same latent space (i.e., the trained neural networks). This has been validated by our experimental results, which will be introduced below. 

\begin{figure}[!thb]\centering
\includegraphics[width=\linewidth]{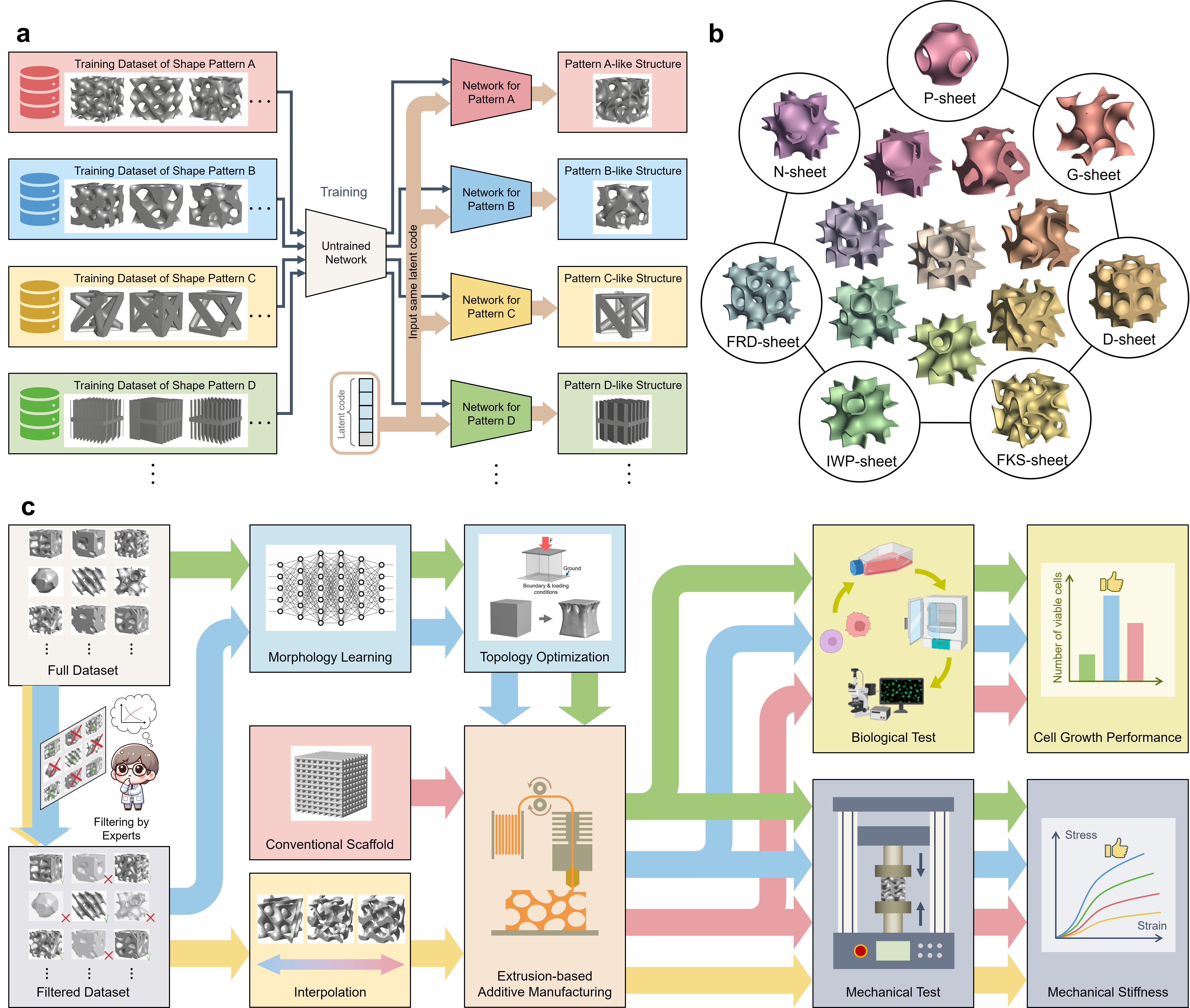}
\caption{\textbf{Achieving simultaneously optimized mechanical and biological performance.} 
\textbf{a} Neural SDF-based morphology learning captures and embeds structural patterns from different datasets into network coefficients, enabling the conversion of latent codes into shapes with varying patterns represented by implicit SDF. 
\textbf{b} Seven distinct types of Triply Periodic Minimal Surfaces (TPMS) are used as the shape basis in our microstructure dataset for morphology learning, where each TPMS type is represented by an implicit basis function -- allowing for the generation of intermediate shapes through interpolation. 
\textbf{c} A dataset filtered by a tissue engineering expert is used to generate a latent space of shape patterns that demonstrate good biological performance (the routine highlighted in blue arrows), where the latent space is then applied in a latent code-based TO process to create structures with enhanced mechanical stiffness. As an alternative (displayed in the green routine), the latent space generated from an unfiltered (complete) dataset tends to result in structures with comparatively poorer biological performance. Random exploration within the latent space of biologically favorable shape patterns, without the guidance of finite element analysis (FEA) driven TO, may lead to structures with significantly weaker mechanical stiffness (as shown in the yellow routine). 
}\label{fig:IdeaOverview}
\end{figure}

Topology Optimization (TO) is a powerful structural design tool aimed at finding the optimal material distribution to maximize structural performance while satisfying specified load conditions and constraints~\cite{bendsoe2013topology,aage2017giga}. Over the past few decades, TO has gained significant attention and has become a prominent research topic, with widespread applications in fields such as solid mechanics~\cite{allaire2004structural,wang2003level}, fluid dynamics~\cite{borrvall2003topology,zhou2008variational}, thermal dynamics~\cite{ha2005topological,yamada2011level,zhuang2007level}, aerospace~\cite{zhu2016topology,guanghui2020aerospace}, automotive~\cite{jankovics2019customization}, and architecture~\cite{jewett2019topology}. Numerous TO methods have been developed across these domains, with some of the most popular and representative being the {Solid Isotropic Material with Penalization} (SIMP)~\cite{bendsoe2013topology}, {Evolutionary Structural Optimization} (ESO)~\cite{xie1993simple}, level-set based~\cite{wang2003level}, and {Moving Morphable Components} (MMC)~\cite{guo2014doing}. TO methods have been employed to optimize micro-structures (also called metamaterials in some literature)~\cite{garner2019compatibility,liu2024ultrastiff}.
While these methods are interrelated, each has distinct characteristics. However, all require solving the static equilibrium equation for intermediate solutions in each iteration, which prevents the incorporation of biological performance during structural shape optimization. 

With the advancement of Artificial Intelligence (AI) techniques, a variety of AI-based TO methods have been developed to either reduce or completely replace the need for iterative operations~\cite{woldseth2022use}. These methods can be broadly categorized into two types: 1) supervised approaches~\cite{abueidda2020topology,yu2019deep}, which primarily rely on image processing pipelines but often suffer from limited generalization capabilities, and 2) unsupervised approaches~\cite{chandrasekhar2021tounn,deng2021parametric,elingaard2022homogenization}, where AI solvers optimize objectives by minimizing loss functions using the auto-differentiation capabilities of neural network-based computing. 
However, because gradient-based optimization necessitates explicitly defining optimization objectives as loss functions, directly incorporating biological performance into the optimization process remains a significant challenge. 
With the advancement of AI techniques, a variety of AI-based TO methods have been developed to either reduce or completely replace the need for iterative operations~\cite{woldseth2022use}.

In this work, we propose a morphology learning-based approach to enable the simultaneous optimization of mechanical and biological performance for microstructures in scaffolds. Cell proliferation is selected to study due to its ease of quantitative evaluation, making it ideal for comparative analysis~\cite{van20223d,jin2024precision}. To preserve the morphological patterns of microstructures while exploring their shape variations, we utilize a Neural SDF-based representation to learn these patterns from a dataset of 3D microstructures (as depicted in Fig.~\ref{fig:IdeaOverview}a). This approach encodes the learned morphology as a network that maps between a latent code and the signed distance function describing the shape of a microstructure. As illustrated in Fig.~\ref{fig:IdeaOverview}a, networks trained on datasets with distinct shape patterns can retain these patterns while navigating the shape space by altering the latent code. Specifically, when the latent space is learned from a filtered dataset containing only microstructures with good biological performance, the likelihood of generating shapes with similar biological characteristics increases when varying the latent code values.

Mechanical stiffness of structures can be maximized by analytically formulating compliance energy as the objective function for optimization~\cite{bendsoe2013topology,aage2017giga}. 
Unlike conventional TO techniques that use the density distribution of the entire design domain as optimization variables, our approach parameterizes structures known for good biological performance by using neural network, allowing for the exploration of shape variations while maximizing stiffness and achieving the required volume fraction. Specifically, we employ latent codes derived from a shape space learned from the filtered dataset as the design variables in TO. This strategy helps preserve favorable biological performance while optimizing mechanical stiffness. Consequently, the microstructures generated by our approach achieve simultaneous optimization of both mechanical stiffness and cell growth. As illustrated in Fig.~\ref{fig:IdeaOverview}c, biological performance is ensured by using the filtered dataset, while mechanical stiffness is optimized through the integration of a neural network-based TO pipeline. As a byproduct, the complex sensitivity analysis required in conventional TO is replaced by the auto-differentiation capabilities of the neural network-based computational pipeline. This offers researchers a powerful tool to define various objective functions, making it easier to handle sensitivity analyses that were previously challenging.

%% file: results.tex
\section*{Results}
\subsection*{Dataset generation and filtering}
To perform morphology learning and generate a latent space for structures with `good' biological performance, we first prepare a dataset of 3D microstructures. Triply Periodic Minimal Surfaces (TPMS) have recently gained significant research attention due to their potential to exhibit not only excellent mechanical properties but also a high area-to-volume ratio~\cite{yang2018mechanical,feng2022triply}. This high area-to-volume ratio makes TPMS structures particularly suitable for cell culture applications. However, not all TPMS structures excel in both mechanical stiffness and cell growth while achieving both simultaneously is even more challenging. The vastness of the design space motivates the introduction of our morphology learning-based approach, which aims to efficiently explore and optimize these complex structures.

\begin{figure}[t]
\centering
\includegraphics[width=0.9\linewidth]{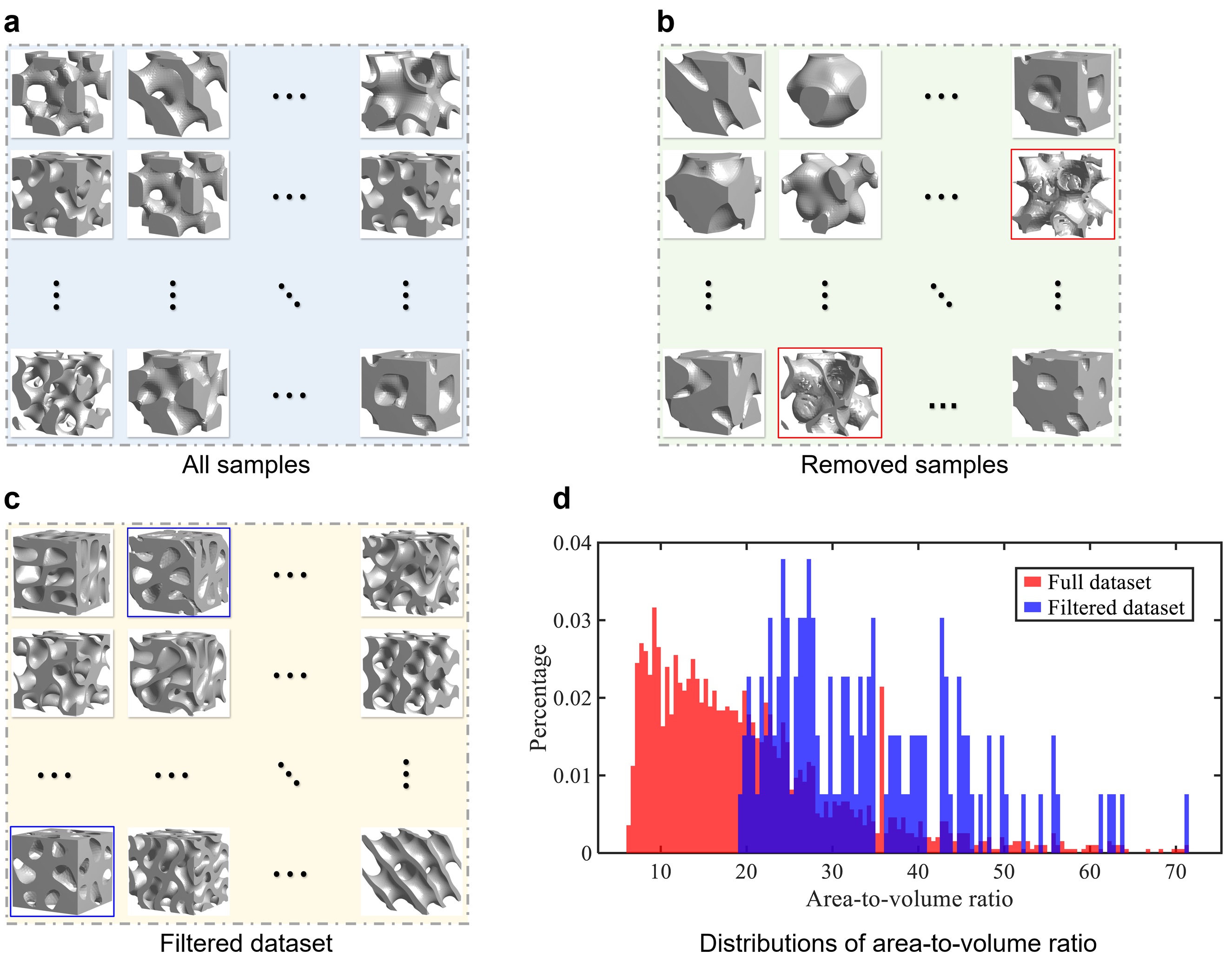}
\caption{\textbf{Dataset filtering.} \textbf{a}~Examples of microstructures in the full dataset (1,960 models in total); \textbf{b}~Examples of removed samples (1,828 models) -- two models with high area-to-volume ratio are highlighted; \textbf{c}~Examples in the filtered dataset (with 132 models) -- two models with low area-to-volume ratio are highlighted; \textbf{d}~Chart of area-to-volume distributions for different datasets -- i.e., the full vs. the filtered datasets, where the vertical axis gives the percentage of samples. 
}\label{fig:dataset}
\end{figure}

Seven different types of TPMS as shown in Fig.~\ref{fig:IdeaOverview}b were used to generate a dataset of microstructures for morphology learning, where each type of TPMS is represented as an implicit basis function. Each sample microstructure in the dataset is generated within the spatial range of a unit cube by linearly blending these seven TPMS functions using weights $\{w_i\}_{i=1,\cdots,7}$, where $w_i \in [0,1]$ and $\sum_i w_i \equiv 1$.  To generate the models, each $w_i$ was uniformly sampled with a step size of 0.05, and then random selections were made from these samples. The final TPMS structure was determined by extracting the isosurface~\cite{ding2021stl}. Additionally, to achieve varying wall thicknesses, the isovalues of the TPMS were randomly sampled within their valid interval~\cite{khaleghi2021directional}. This process resulted in a dataset composed of $1,960$ microstructures, with examples shown in Fig.~\ref{fig:dataset}a. The surface areas of the models in this dataset change in the range $[2.371,13.17]$ with the range of volumes as $[0.1010, 0.9515]$. The range of area-to-volume ratios for samples in this dataset is $[6.159,72.18]$. 

After consulting with tissue engineering experts, we discovered that not all TPMS structures in our dataset are suitable for cell culture applications. Cell culture specialists were engaged to filter the dataset of $1,960$ models. Each structure was evaluated for its suitability by rotating and inspecting its internal features through slicing. As a result, $1,828$ models are removed (see Fig.~\ref{fig:dataset}b for removed examples), leaving $132$ models in the filtered dataset $\mathcal{D}$ (see Fig.~\ref{fig:dataset}c for retained examples). This filtering revealed an interesting trend: models with a smaller area-to-volume ratio were more likely to be excluded due to poorer porosity. This observation is supported by the sample distribution charts shown in Fig.~\ref{fig:dataset}d, which indicate that the filtered dataset has a higher area-to-volume ratio compared to the original dataset. However, area-to-volume ratio alone cannot serve as a simple criterion for evaluating cell growth performance. For instance, the models highlighted by red frames in Fig.~\ref{fig:dataset}b, which have relatively larger area-to-volume ratios (e.g., $72.13$ for the one in the second row and $60.54$ for the one at the bottom row), were still removed. 
This exclusion was due to the structures' weak connection to the surrounding host tissue and their low interconnectivity within the scaffold, making it difficult to ensure adequate permeability during tissue regeneration. 
Conversely, some models with smaller area-to-volume ratios (e.g., $19.64$ for the one in the first row and $19.12$ for the one in the bottom row, highlighted by blue frames in Fig.~\ref{fig:dataset}c) were retained because of their relatively larger contact area to link with the local host tissue at the scaffolds external surface 
while maintaining the inter-connectivity. In summary, the decision to remove or retain samples was based on multiple complex factors, with a conservative strategy applied during filtering. Structures with uncertain biological performance were excluded from the dataset. Consequently, the filtered dataset $\mathcal{D}$ contains 132 models. 


\begin{figure}[t]
\centering
\includegraphics[width=0.8\linewidth]{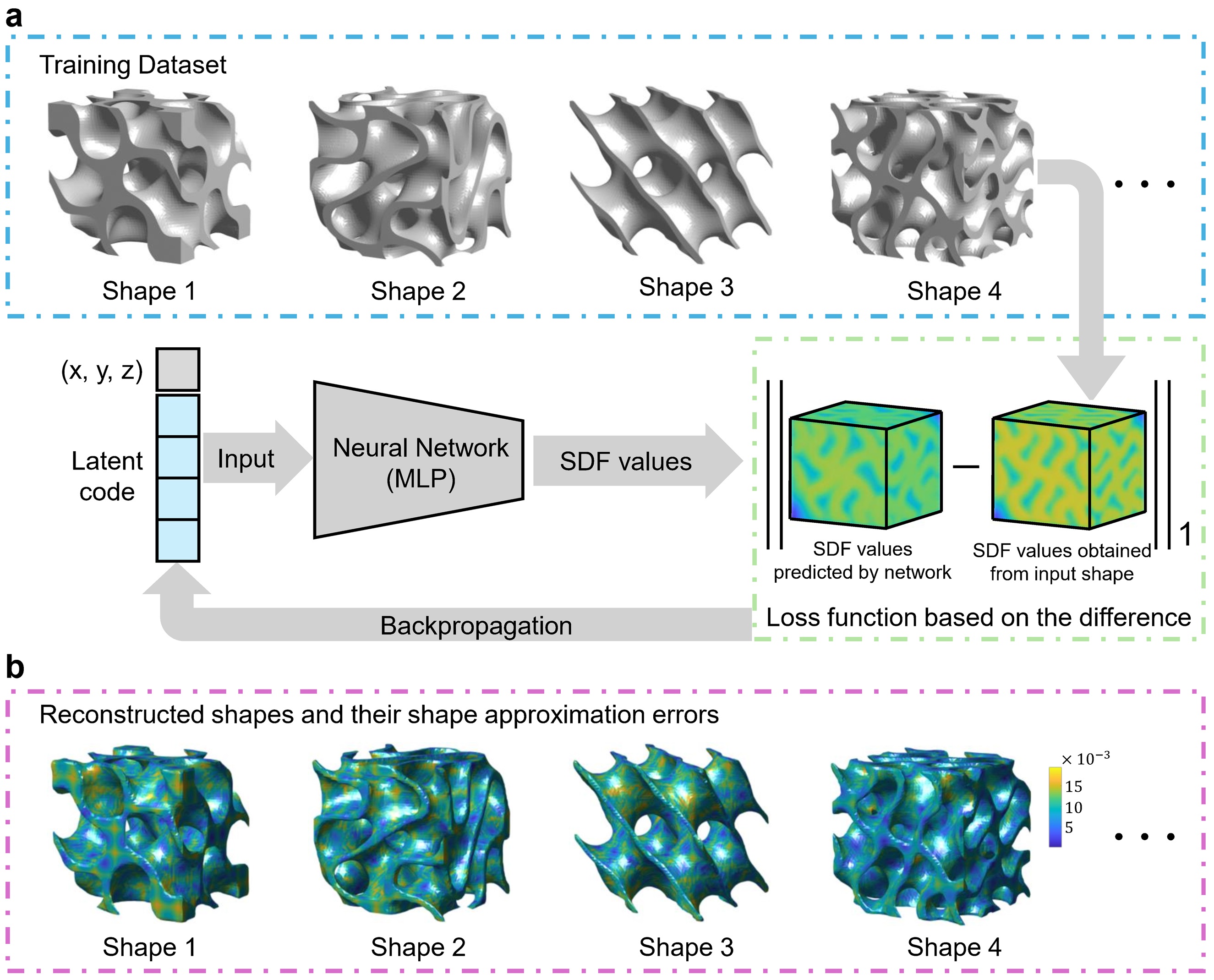}
\caption{\textbf{Training of neural network $f_\theta(\cdot)$ for morphology learning.} \textbf{a}~The coefficients $\theta$ of the network (shared for all models in $\mathcal{D}$) and the latent code $\mathbf{z}_i$ for each model $H_i \in \mathcal{D}$ are employed as unknown variables to be optimized during the training process, where the loss of training is the difference between $f_\theta(\cdot)$ and the SDF values of all models in $\mathcal{D}$. \textbf{b}~The models generated by the neural network $f_\theta(\cdot)$ and the latent codes obtained during training, where the shape approximation errors are given in the color map indicating the high accuracy of reconstruction.
}\label{fig:learning}
\end{figure}

\subsection*{Model architecture, training and performance}
Inspired by the concept of Signed Distance Functions (SDF) in implicit solid modeling, any 3D solid model $H$ can be mathematically described by a function $s(\mathbf{x})$, where the returned value represents the signed distance from a query point $\mathbf{x} \in \mathbb{R}^3$ to the boundary surface of $H$. This value is negative for points inside the model and positive for those outside. The boundary surface of $H$ is defined by the set of points satisfying $s(\mathbf{x})=0$. The core idea of morphology learning in our work is to train a neural network $f_\theta(\mathbf{x},\mathbf{z})$, where a $d$-dimensional vector $\mathbf{z} \in \mathbb{R}^d$ serves as the latent code that controls shape variation. Specifically, for a fixed latent code $\mathbf{z}^i$ corresponding to a solid model $H_i$ ($i=1,2,\cdots$), the network $f_\theta(\mathbf{x},\mathbf{z}^i)$ is expected to accurately represent the SDF of the model $H_i$. The training objective is to ensure that this relationship holds true for every model within the training dataset. 

In our experiment, we utilize a Multi-Layer Perceptron (MLP) neural network for training, consisting of 5 layers, each with 512 neurons, and employing ReLU activation functions to capture nonlinear shape variations. Training is conducted on the filtered dataset $\mathcal{D}$. For every model $H_i$ in $\mathcal{D}$, we first compute its accurate SDF by a geometric computing approach~\cite{baerentzen2005signed}. The loss of training is defined as the $L_1$ norm of the difference between the accurate SDF value and the SDF value approximated by the network $f_\theta(\cdot)$ in the computational domain (i.e., the spatial range of a unit cube). The $L_2$ norm of the latent code $\mathbf{z}$, $\| \mathbf{z} \|^2$, is employed as a regularization term for training. The latent code is set to a dimension of 128. The network is trained over 3000 epochs, starting with an initial learning rate of $5\times10^{-4}$, which is halved every 500 epochs. An illustration of the morphology learning process can be found in Fig.~\ref{fig:learning}a. 

After training, we obtain a differentiable function space representation $f_{\theta}(\cdot)$ that also encodes the shape pattern of structures embedded in the training dataset. When using the latent code obtained via training, the shape of a structure can be reconstructed in high accuracy -- see the shape approximation error maps given in Fig.~\ref{fig:learning}b. When exploring the function space $f_{\theta}(\cdot)$ by randomly changing the values of latent code in the range $[-1.0,1.0]$, the shape pattern embedded in the training dataset can be well preserved. More details of the morphology exploration will be discussed below.

\begin{figure}[!thb]
\centering
\includegraphics[width=1\linewidth]{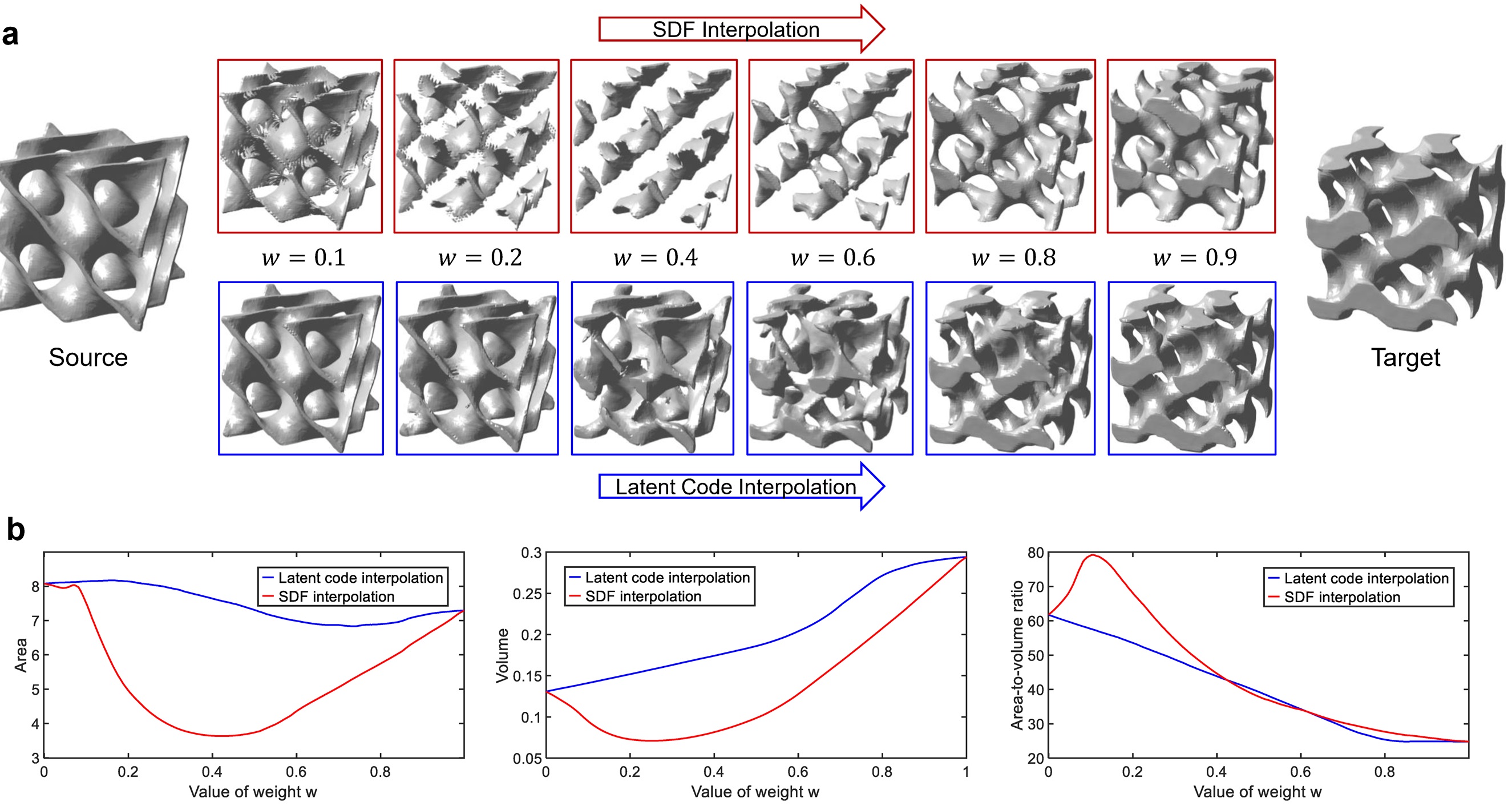}
\caption{\textbf{Importance of latent code-based morphology exploration demonstrated by interpolation between two structures.} 
\textbf{a}~Structures generated by direct SDF interpolation (top row) versus latent code-based interpolation (bottom row). The results from latent code-based interpolation show a better preservation of the shape patterns from the source and target structures.
\textbf{b}~Changes in surface area, volume, and area-to-volume ratio are compared between the two interpolation methods. The latent code-based interpolation exhibits much smoother transitions, indicating a more consistent preservation of shape characteristics.
%
}\label{fig:interp_ST}
\end{figure}

\subsection*{Morphology exploration}
The advantages of using latent codes derived from morphology learning will be explored in this sub-section. First, when training the neural SDF network with datasets containing models of different shape patterns, random variations in the latent code consistently produce shapes that adhere to the same pattern. As illustrated in Fig.~\ref{fig:IdeaOverview}a, a network trained on TPMS structures will not generate a truss-like structure when the latent code is altered. Conversely, a TPMS structure cannot be produced by a network trained solely on a dataset of truss-like structures. This pattern consistency ensures that the morphological characteristics inherent to the original dataset are preserved, even when exploring new shape variations. Moreover, this approach significantly narrows the search space, allowing for more targeted exploration of design variations that maintain desirable properties, such as mechanical strength or biological compatibility.

Experiments on shape interpolation were conducted to evaluate the representational capacity of the latent code. Given two structures, $H_a$ and $H_b$, with their respective SDFs denoted as $s_a(\mathbf{x})$ and $s_b(\mathbf{x})$, the linear interpolation between these two SDFs can be directly computed as
\begin{equation}\label{eq:sdf_int}
    s(w,\mathbf{x}) = (1-w) s_a(\mathbf{x}) + w s_b(\mathbf{x}),
\end{equation}
where different values of $w$ yield various intermediate shapes (see the top row of Fig.~\ref{fig:interp_ST}a). 
However, this straightforward interpolation often fails to naturally preserve the original shape patterns of $H_a$ and $H_b$. In contrast, when $H_a$ and $H_b$ are represented by their respective learned latent codes, $\mathbf{z}_a$ and $\mathbf{z}_b$, the interpolation in latent space produces shapes as SDF defined by
\begin{equation}\label{eq:lc_int}
    \Tilde{s}(w,\mathbf{x}) = f_\theta(\mathbf{x}, (1-w)\mathbf{z}_a + w \mathbf{z}_b).
\end{equation}
As shown in the bottom row of Fig.~\ref{fig:interp_ST}a, this latent code-based interpolation effectively preserves the distinct shape patterns of $H_a$ and $H_b$ across the interpolation range, demonstrating the robustness of capturing shape patterns. 
In addition to comparing the shapes, we also quantitatively analyzed the changes in surface area, volume, and the area-to-volume ratio between SDF interpolation and latent code-based interpolation. The results are plotted as curves in Fig.~\ref{fig:interp_ST}b. It is evident that direct SDF interpolation results in significant fluctuations in all these metrics, primarily due to the drastic changes in topology during the middle range of interpolation. In contrast, latent code-based interpolation produces much smoother transitions, reflecting a more consistent preservation of shape characteristics.

\begin{figure}[t]
\centering
\includegraphics[width=\linewidth]{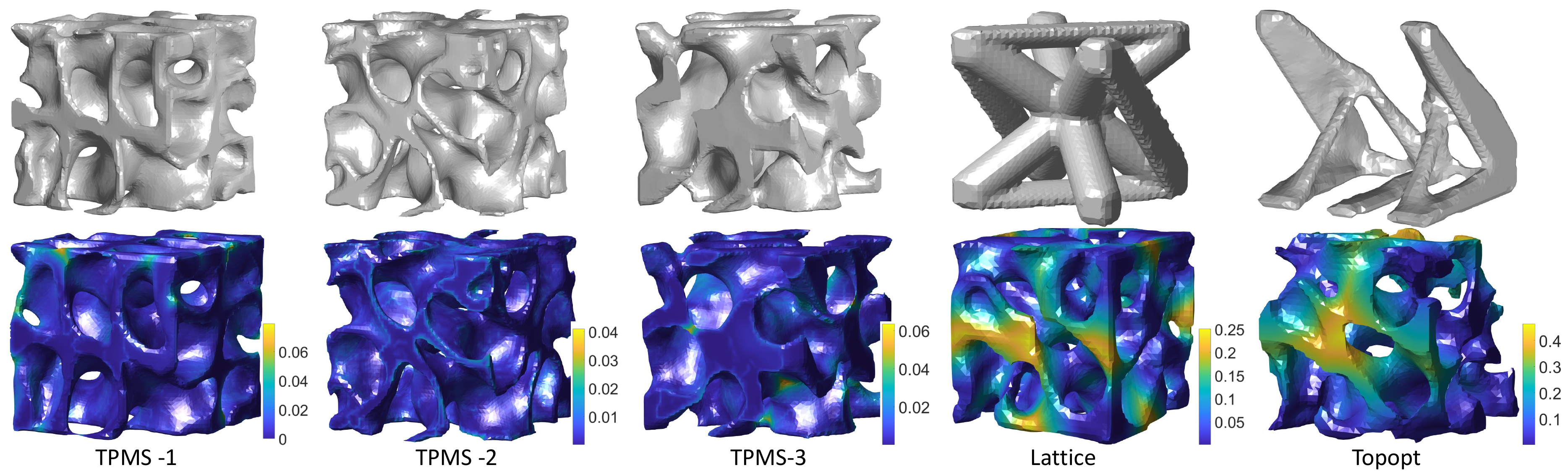}
\caption{\revise{}{\textbf{Shape approximation by the latent code-based representation.} The neural network $f_{\theta}(\mathbf{x},\mathbf{z})$ learned from a TPMS dataset is employed to approximate: an unseen TPMS model, a lattice structure, and a model of intersected plates in the top row. The geometric approximation errors are visualized by color maps on the surfaces of the reconstructed models shown in the bottom row.}
}\label{fig:LatentCodeRepApproximation}
\end{figure}

\revise{}{We conducted an additional study to demonstrate the generalization capability of our latent code-based representation obtained through morphology learning. Specifically, as illustrated in Fig. \ref{fig:LatentCodeRepApproximation}, we utilized a neural network $f_{\theta}(\mathbf{x},\mathbf{z})$ trained on a TPMS dataset to approximate various unseen models. For a new model represented by the SDF function $s^w(\mathbf{x})$, its approximation is achieved by solving}
\begin{equation}
\revise{}{\mathbf{z}_w = \arg \min_{\mathbf{z}} \sum_{\mathbf{x}} | f_{\theta}(\mathbf{x},\mathbf{z}) - s^w(\mathbf{x})|.}
\end{equation}
\revise{}{The resultant latent code $\mathbf{z}_w$ is determined through backpropagation, as illustrated in Fig. \ref{fig:learning}a, while keeping the network coefficients unchanged. The geometric approximation error is evaluated as the difference between $s^w(\mathbf{x})$ and $f_{\theta}(\mathbf{x},\mathbf{z}_w)$ at every surface point $\mathbf{x}$ where $s^w(\mathbf{x})=0$. Notably, the approximation errors for the unseen TPMS model are significantly smaller compared to the other two unseen models. This is because the shape patterns learned and stored in $f_{\theta}(\cdot)$ are derived from the TPMS dataset, which exhibits substantial variations compared to the lattice and plate models.
}

\begin{figure}[t]
\centering
\includegraphics[width=\linewidth]{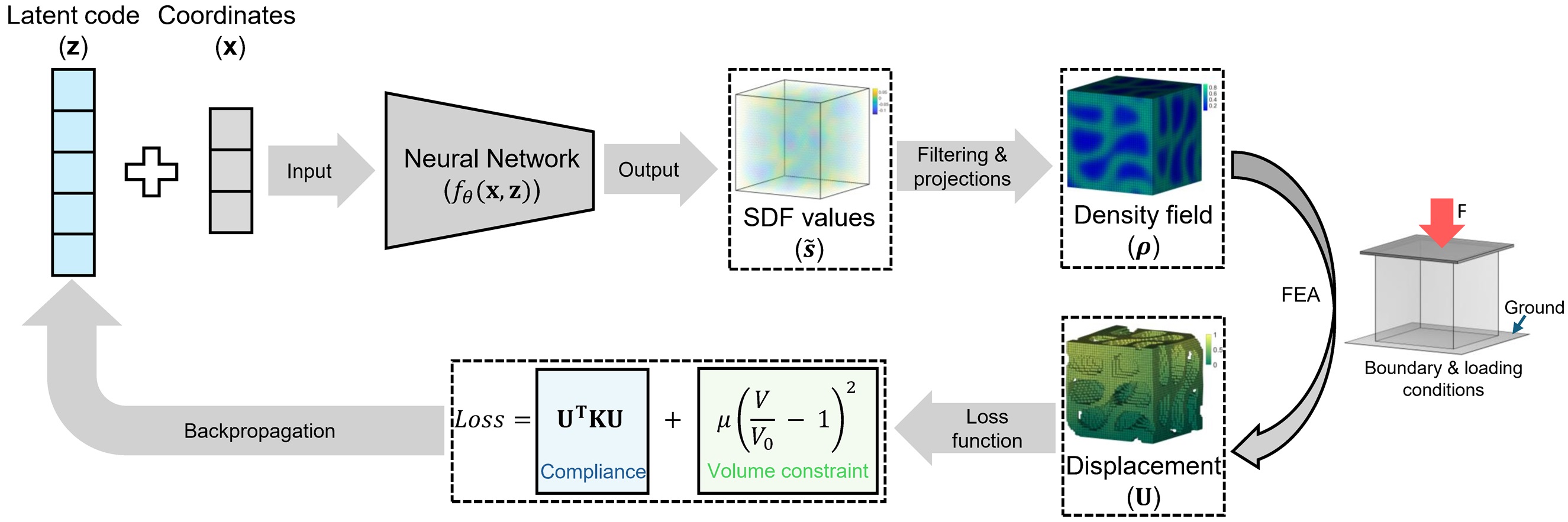}
\caption{\textbf{Computational pipeline of our pattern-preserving TO.} The optimized structure can be obtained by updating the values of the latent code $\mathbf{z}$ through the backpropagation of the neural network-based self-learning.
}\label{fig:TOBackpropagation}
\end{figure}

%
%
%
%

\begin{figure}[htb]
\centering
\includegraphics[width=0.9\linewidth]{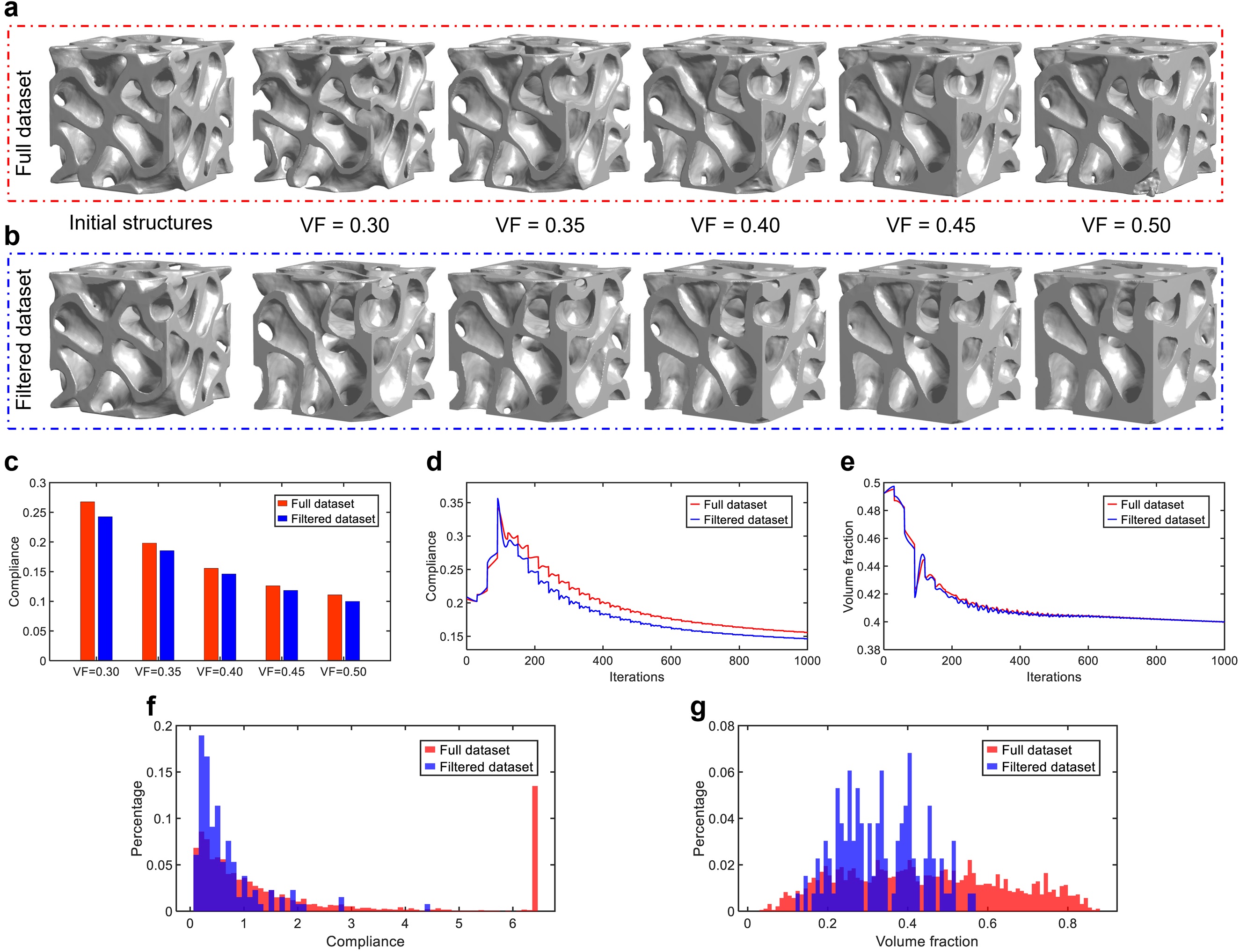}
\caption{\textbf{Structures with compliance minimized by using latent-codes learned from different datasets.} 
\textbf{a}~A structure is randomly selected from the filtered dataset, which also exists in the full dataset, and its latent codes from both datasets are obtained to serve as the initial values for optimization. 
\textbf{b}~The structures optimized with different target volume fractions (VF) by using the network learned from the full dataset (top row) and the network learned from filtered dataset (bottom row)).
\textbf{c}~Bar-chart for comparing the compliance of resultant structures. 
\textbf{d}~Convergence curves of compliance for the structure optimized with VF = 0.4 by using the full dataset vs the filter dataset. 
\textbf{e}~Convergence curves of volume fraction for the latent code based structure optimization with VF = 0.4.
\textbf{f}~The distribution of compliance for structures in both datasets.
\textbf{g}~The distribution of volume fraction for structures in both datasets.
}\label{fig:complianceOptm_fromDiffDatasets}
\end{figure}

\subsection*{Structures optimization}
Using the latent space derived from morphology learning, we optimize structures to enhance their mechanical performance while preserving the original shape patterns. The workflow of our pattern-preserving structural optimization method is illustrated in Fig.~\ref{fig:TOBackpropagation}. The design domain is tessellated into a grid of $40\times40\times40$ voxels, which serve as elements for finite element analysis (FEA) under compression loading. As shown on the right of Fig.~\ref{fig:TOBackpropagation}, the points at the base of the design domain are fixed, and the top surface is subjected to distributed forces generated by placing a thin shell at the top of the design domain. This setup, including the boundary conditions and loading specifications, is consistently used across all experiments in this study.

Given a latent code $\mathbf{z}$, a corresponding SDF can be obtained from the neural network $f_{\theta}$ trained during morphology learning. The SDF values for all finite elements are computed using the central points $\mathbf{x}_e$ as $f_{\theta}(\mathbf{z},\mathbf{x}_e)$. Ideally, elements with $f_{\theta}(\mathbf{z},\mathbf{x}_e) \leq 0$ are considered solid, while those with $f_{\theta}(\mathbf{z},\mathbf{x}_e) > 0$ are treated as void. However, a direct binary classification like this would make the problem non-differentiable. Instead, the SDF values are filtered and projected to produce a continuous density field that represents the distribution of material within the design domain, where a density of 1 indicates a solid element and a density of 0 indicates a void. During optimization, material properties are applied with a Young's modulus of $1.0$ for solid elements and a Poisson's ratio of $0.3$ for all elements. The Young's modulus for an element with density $\rho \in [0.0,1.0]$ is then obtained through $p$-norm weighted interpolation between $E_{\max}=1.0$ and $E_{\min} = 10^{-3}$ by the density $\rho_e$ of element $e$. 

The stiffness matrix for each element is constructed to assemble the global system stiffness matrix, $\mathbf{K}$, allowing the displacements $\mathbf{U}$ of all nodes to be determined through FEA by solving the equation $\mathbf{K} \mathbf{U} = \mathbf{F}$. Mechanical stiffness can be maximized by minimizing the compliance~\cite{sigmund200199}, which is evaluated by $\mathbf{U}^T \mathbf{K} \mathbf{U}$. Additionally, the structure optimization must satisfy a volume constraint
$V \leq V_0$ with $V_0$ representing the maximum allowable volume for the optimized structure. The structures, which retain the shape pattern similar to those in the dataset used for morphology learning, are obtained by updating the latent code $\mathbf{z}$ to minimize a loss function that combines the compliance energy and the volume constraint, represented as $(V/V_0 -1)^2$. The latent code is iteratively updated via backpropagation driven by this loss function, as illustrated in Fig.~\ref{fig:TOBackpropagation}.

Using this structural optimization framework, we can generate structures with high mechanical stiffness that do not exist in the original dataset. Moreover, these structures will retain the common patterns inherent in the dataset, as they are encoded by the neural network that translates latent codes into the SDFs of the models. 

\subsection*{Mechanical stiffness}
In the first mechanical experiment, we evaluate the mechanical stiffness of the structures optimized by using the latent spaces (i.e., the networks) learned from both the full dataset and the filtered dataset. A structure is first randomly selected from the filtered dataset, which also exists in the full dataset (see the left-most one in Fig.~\ref{fig:complianceOptm_fromDiffDatasets}a\&b). The latent codes of this structures are also obtained from the two datasets to serve as the initial values for the computation of structure optimization. The objective function combines structural compliance and volume fraction, with five volume fractions used in the optimization as: 0.3, 0.35, 0.4, 0.45, and 0.5. The optimized structures are shown in Fig.~\ref{fig:complianceOptm_fromDiffDatasets}a\&b. 
The compliance values of the optimized structures are presented in Fig.~\ref{fig:complianceOptm_fromDiffDatasets}c, where lower values indicate greater structural stiffness. Interestingly, the structures optimized within the shape space learned from the filtered dataset exhibit greater stiffness than those optimized within the space from the full dataset. This is due to the full dataset containing many structures with weak stiffness under the loading conditions defined in Fig.~\ref{fig:TOBackpropagation}. As a result, the average performance of structures from the full dataset is slightly reduced compared to that of the filtered dataset. 
This experiment also highlights that 
careful selection of structures to include into the dataset is crucial. 
%
The convergence curves of compliance and volume fraction for the structure optimized using VF = 0.4 in the filtered dataset are shown in Fig.~\ref{fig:complianceOptm_fromDiffDatasets}d and ~\ref{fig:complianceOptm_fromDiffDatasets}e. The bumps in these curves are caused by dynamic changes in the projection parameter. The volume fraction of the optimized structure converges well to 0.4, and compliance is effectively minimized. 

\begin{figure}[t]
\centering
\includegraphics[width=1\linewidth]{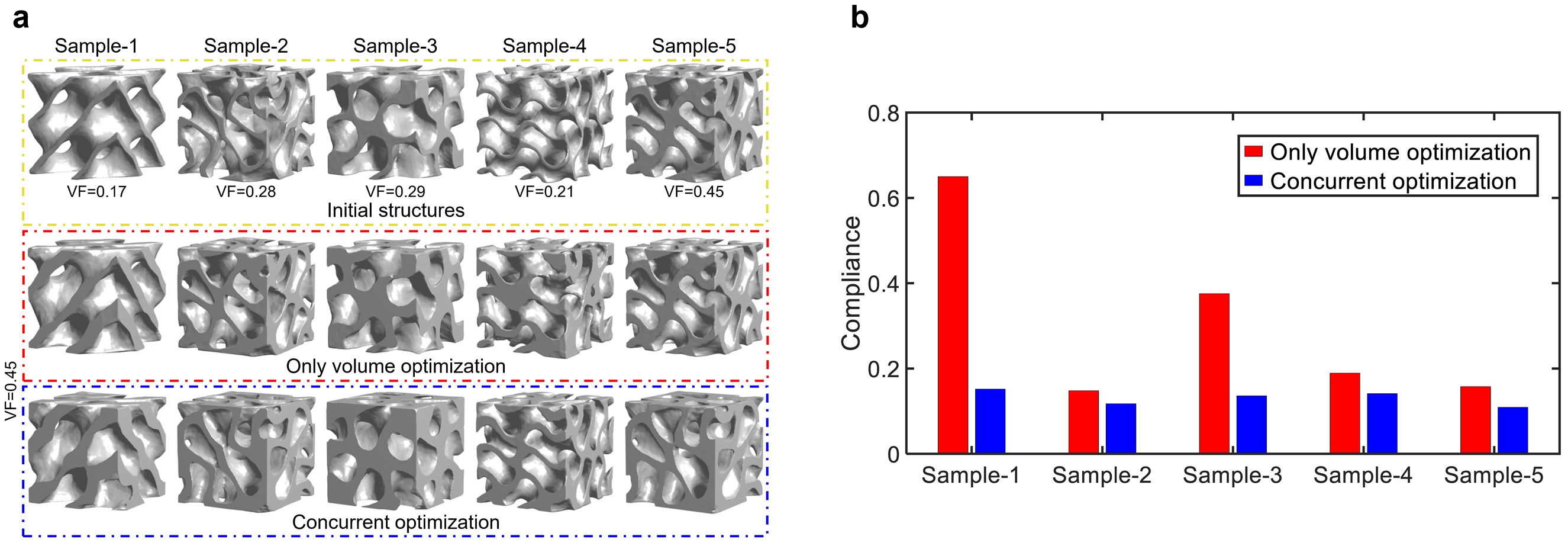}
\caption{\textbf{Comparison of the mechanical stiffness of structures obtained by optimizing only the volume vs. those obtained by concurrently optimizing both the compliance and the volume.} 
\textbf{a}~Five samples are randomly selected from the filtered dataset, which are shown in the top row with their volume fractions (VF values). They are optimized to a target volume fraction as 0.45. The second row shows the structures obtained by optimizing only the volume, while the bottom row shows the results by optimizing both the volume and the structural compliance (therefore the stiffness).
\textbf{b}~Comparison of the compliance values on resultant structures.}\label{fig:SV_or_V}
\end{figure}

\begin{figure}[t]
\centering
\includegraphics[width=1\linewidth]{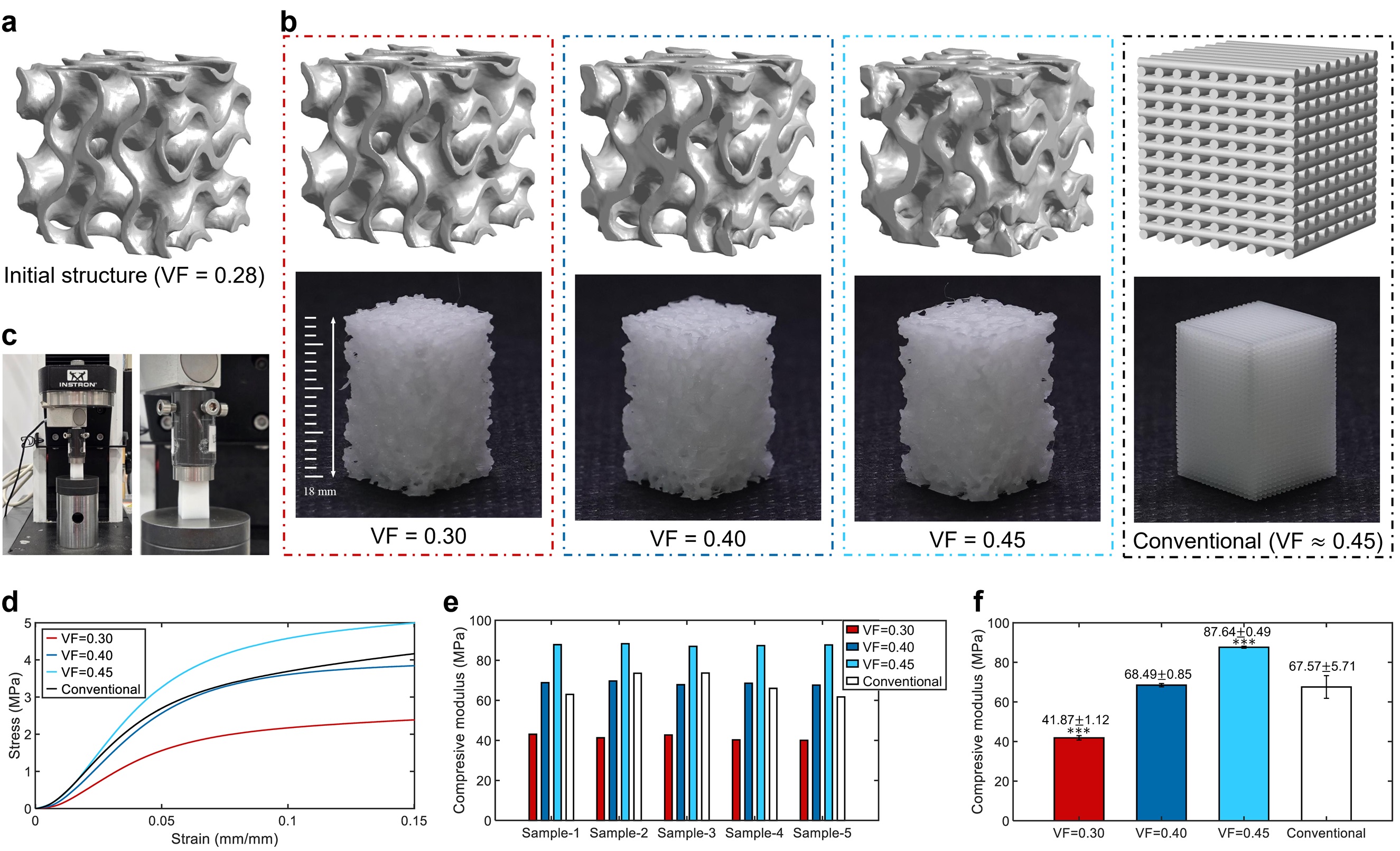}
\caption{\textbf{Mechanical compression tests of structures optimized by using the latent-code learned from the filter dataset vs. the conventional scaffold.} 
\textbf{a}~An initial structure with a volume fraction of 0.28.
\textbf{b}~Three compliance optimized structures are obtained with different target volume fractions (VF). The right-most shows the conventional scaffold with its volume fraction being about 0.45. The scaffold specimens with dimensions $12\mathrm{mm} \times 12\mathrm{mm} \times 18\mathrm{mm}$ 
that are 3D printed according to different structures shown in the first row.
\textbf{c}~The hardware setup employed in our mechanical compression tests to verify the mechanical stiffness of structures generated by our method via measuring the compression modulus. 
%
\textbf{d}~Experimental results as stress-strain curves. 
\textbf{e}~Bar-charts of the compression modulus values of each sample.
\textbf{f}~Bar-chart of the compression modulus values with standard derivation tested on 5 samples for each structure. \revise{}{The significance levels are set as * p < 0.05, ** p < 0.01, and *** p < 0.001.}
}\label{fig:different_volume}
\end{figure}

In the second experiment for mechanical stiffness, we study the effectiveness of our optimizer for stiffness improvement. Two different optimizations are conducted: (i) only considering the volume requirement and (ii) concurrently considering both the volume and the compliance objectives. Five structures are randomly sampled from the filtered dataset by using their latent codes to initialize the optimization process with these initial structures are shown in the top row of Fig.~\ref{fig:SV_or_V}a. When setting the target volume fraction at 0.45, the resultant structures by only considering volume are as shown in the middle row of Fig.~\ref{fig:SV_or_V}a. The optimized structures considering both the volume and the compliance are given in the bottom row of Fig.~\ref{fig:SV_or_V}a. Their corresponding values of compliance are visualized in the bar-chart of Fig.~\ref{fig:SV_or_V}b. This experimental test proves that our method can effectively reduce the compliance on resultant structures no matter whether the initial structure has significantly different volume fraction (e.g., Sample-1) or similar volume fraction (e.g., Sample-5). As a consequence, structures with larger stiffness can be consistently obtained by our optimizer. 

The effectiveness of compliance based optimization is also verified by mechanical compression tests. Specifically, we rank the area-to-volume ratios of all structures in the dataset in descending order and randomly select one from the top 10\% of the filtered dataset as the initial structure of optimization. Three structures with different volume fractions such as 0.3, 0.4 and 0.45 are generated by using the latent code learned from the filtered dataset -- see Fig.~\ref{fig:different_volume}a. A conventional scaffold widely employed in prior research of bone tissue engineering is also shown at the most-right of Fig.~\ref{fig:different_volume}b with its volume fraction being 0.45. In detail, the conventional scaffolds have a 0$^\circ$/90$^\circ$ lay-down pattern with 330 $\upmu$m fiber diameter, 350 $\upmu$m pore size, and 270 $\upmu$m layer thickness. These parameters were optimized, specifically, based on the typical vertical / horizontal structures to balance both mechanical and biological performance of the scaffolds, meeting the design criteria of the bone scaffold~\cite{eckhart2019covalent,freed2009advanced,lee2019resolution}. 

The computational results of all these structures as their compliance values are: $0.2798$ (VF = 0.3), $0.1859$ (VF = 0.4) and $0.1559$ (VF = 0.45), where the conventional scaffold has its compliance as $0.1747$ -- i.e., less than our result with VF=0.4 but larger compliance than our result with VF=0.45. For the optimized structures and the conventional scaffold, we arrange them periodically within a given cubic domain with dimensions as $12\mathrm{mm} \times 12\mathrm{mm} \times 18\mathrm{mm}$ and fabricate the periodically repeated structures by filament deposition based 3D printing (see the physical specimens as shown in Fig.~\ref{fig:different_volume}b). Uni-axial mechanical compression tests were conducted according to the ASTM standards to evaluate the compressive modulus of these structures by using the hardware as shown in Fig.~\ref{fig:different_volume}c. 

The results of compression tests are given in Fig.~\ref{fig:different_volume}d-e. As can be observed, all scaffolds can match the modulus of human cancellous bone (i.e., 20-500 MPa~\cite{dziaduszewska2021structural, gerhardt2010bioactive}), enabling them to withstand the hydrostatic and pulsatile pressures found in physiological environments while maintaining the necessary pores for cell attachment, proliferation, and differentiation \cite{chung2011design, zerankeshi2022polymer}. Results also show that the compressive modulus increases significantly with the volume fraction (VF), ranging from 42.47 MPa at VF = 0.3 to 87.98 MPa at VF = 0.45, offering considerable design flexibility to match cancellous bone at different anatomical sites. Moreover, optimized scaffolds at VF = 0.4 exhibited a similar compressive modulus to conventional scaffolds but with a lower volume fraction. Conversely, at the same volfrac (VF=0.45), the optimized scaffold demonstrated a compressive modulus (89.98 MPa) that is significantly higher than the conventional scaffold (69.63 MPa), highlighting the effectiveness of the optimization in enhancing mechanical stiffness that is measured as compressive modulus in physical experiments.

\vspace{10pt}

\subsection*{Biological characterization}

\begin{figure}[t]
\centering
\includegraphics[width=1\linewidth]{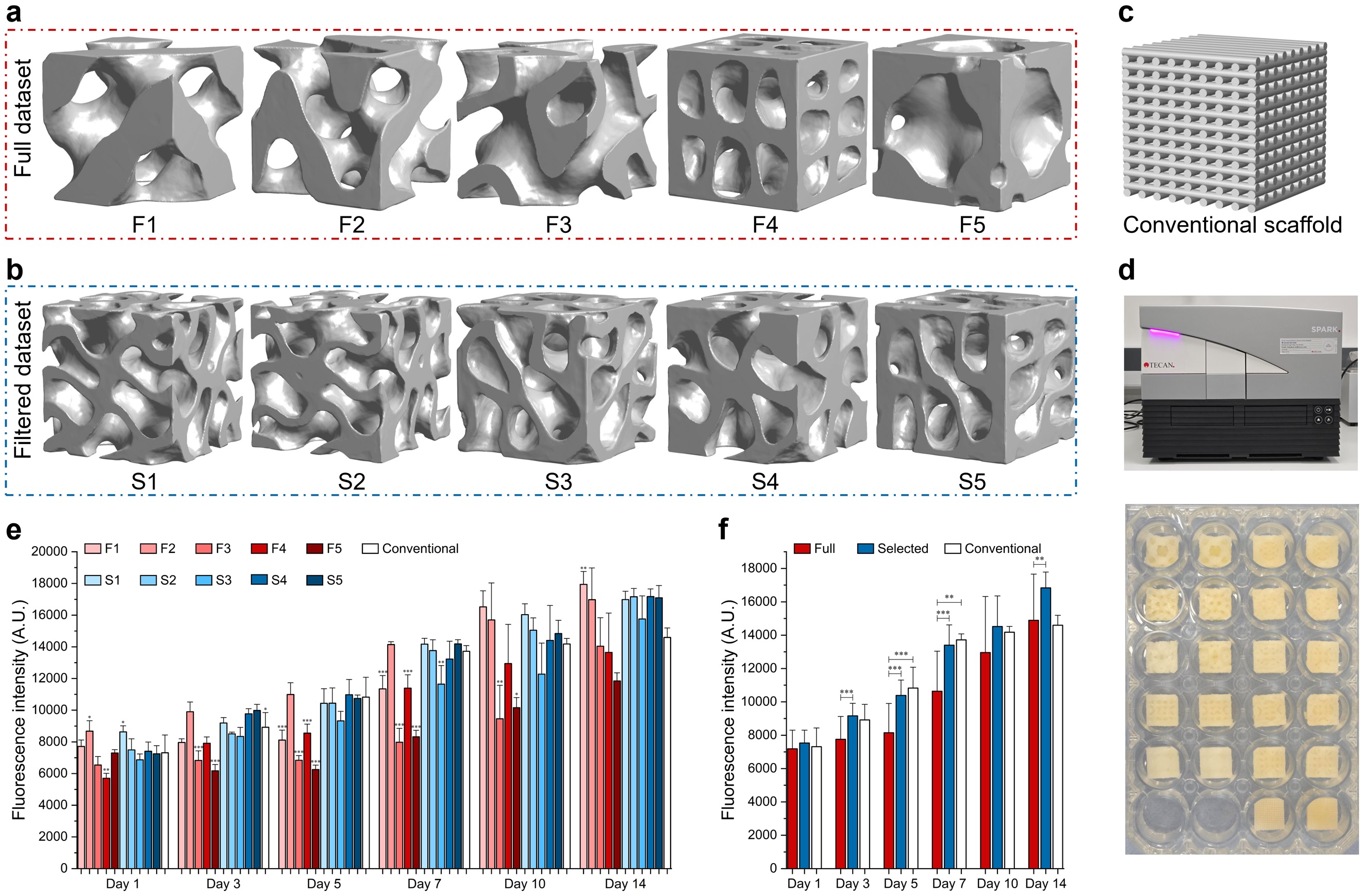}
\caption{\textbf{Biological performance verification for the structures generated by the latent-code learned from the filter dataset.}
\textbf{a}~The structures generated by optimizing samples randomly selected from the full dataset into a volume fraction as 0.45. 
\textbf{b}~The structures generated by optimizing samples randomly selected from the filtered dataset -- all resultant structures have VF=0.45. 
\textbf{c}~The conventional scaffold design taken for comparison purpose. 
\textbf{d}~The Alamar Blue assay was selected to assess and compare the cell proliferation across various structures, evaluating their biological performance. 
\textbf{e}~The performance in hADSC proliferation results\revise{}{, where the measurements are taken on 5 specimens for each group -- i.e., the bar-chart is generated by the mean values with standard derivation}. 
\textbf{f}~The averaged performance across the five structures for each dataset. \revise{}{The significance levels are set as * p < 0.05, ** p < 0.01, and *** p < 0.001.}
}\label{fig:filtering}
\end{figure}

\paragraph{Filtering effectiveness}
To validate the effectiveness of the dataset filtering for the cell culture application, two groups of structures are constructed and their biological performances are compared. One group is constructed by randomly selecting five structures from the full dataset (denoted as F1-F5), while the other group is formed by randomly selecting five structures from the filtered dataset (denoted as S1-S5). After that, the volume of these structures in both groups are optimized to be VF = 0.45, which is the same volume fracture of the conventional scaffold. The optimized structures are shown in Fig.~\ref{fig:filtering}a and Fig.~\ref{fig:filtering}b. The mechanical stiffness of all these structures has not been optimized. Their biological performance is compared with the conventional design of scaffold (see Fig.~\ref{fig:filtering}c) via cell proliferation analysis considering hADSc cell line (Fig.~\ref{fig:filtering}d). The results are shown in Fig.~\ref{fig:filtering}e. 
It can be observed that cell proliferation on the structures generated from the full dataset presented significant variations, reflecting that the full dataset contains both relatively good (e.g., F1 and F2) and undesirable structures (e.g., F3 and F5). In contrast, the structures generated from the filtered dataset showed better cell proliferation results with significantly less fluctuation, indicating a more consistent design set in terms of biological performance. From the combined analysis (as shown in Fig.~\ref{fig:filtering}f), we can observe that the S group presents significant advantage over the F group on all time points except day 1 and day 10. Additionally, although without statistical difference, the cell proliferation behaviour of S group was also higher than the conventional scaffold after day 10, indicating its strong potential for further optimisation.

\begin{figure}[t]
\centering
\includegraphics[width=0.8\linewidth]{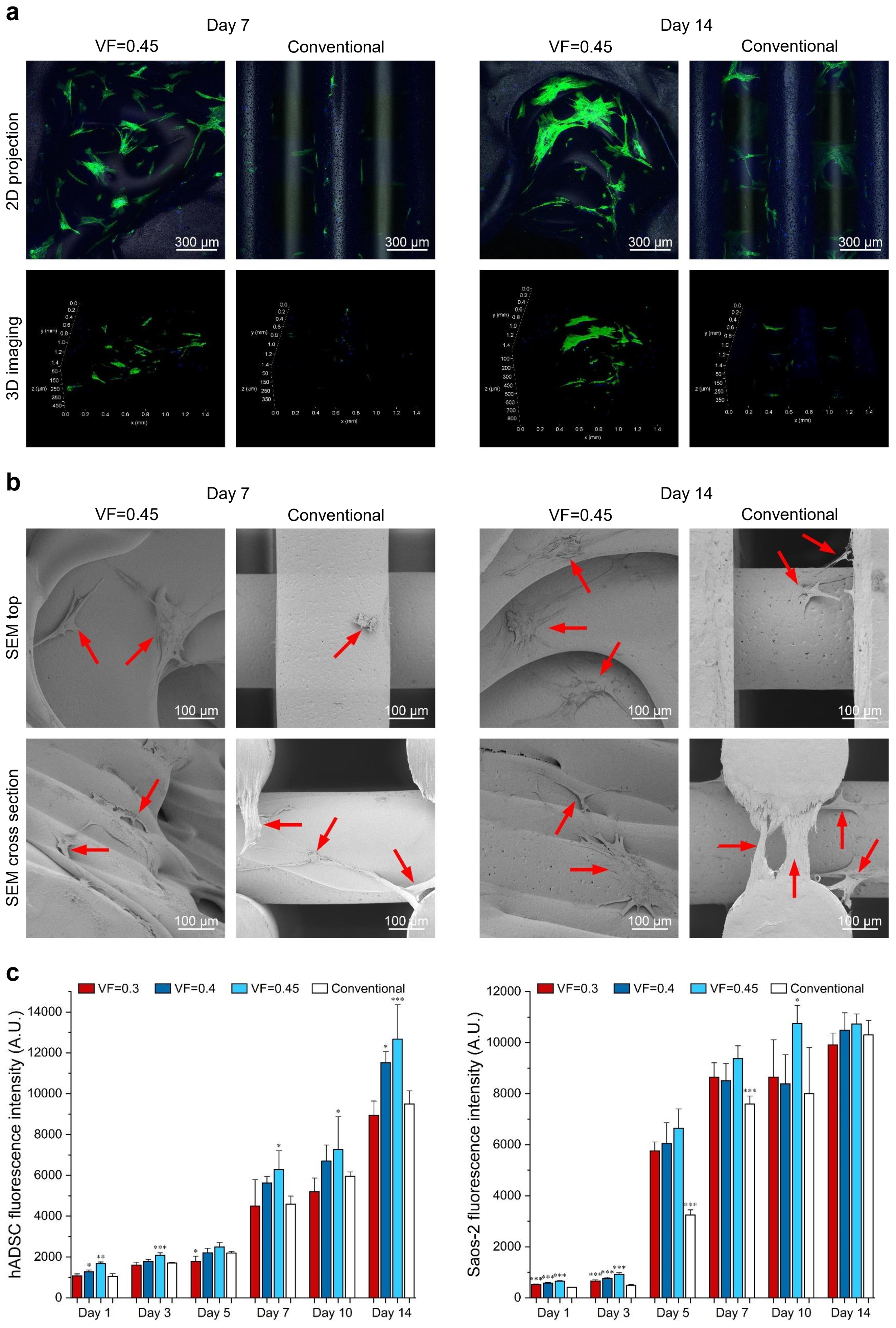}
\caption{\textbf{Bio-images and cell proliferation of different scaffold designs.} \textbf{a} Confocal microscopy images ($\times80$ magnification) and \textbf{b} SEM images ($\times500$ magnification) of hADSC on our optimized scaffold with VF=0.45 and the conventional scaffold at different time points, where red arrows represent the proliferated cells. 
Cell nuclei is stained blue (DAPI) and F-actin is stained green (Alexa Fluor 488) in confocal images. 
\textbf{c} hADSC and \textbf{d} Saos-2 cell viability and proliferation results on scaffolds with different volume fractures. \revise{}{The significance levels are set as * p < 0.05, ** p < 0.01, and *** p < 0.001.}
}\label{fig:Proliferation}
\end{figure}

\paragraph{Effectiveness of concurrent optimization}

Both confocal microscopy (Fig.~\ref{fig:Proliferation}a) and SEM (Fig.~\ref{fig:Proliferation}b) were considered for bioimaging, to illustrate cell morphology and spreading status on scaffolds. As seen in Fig.~\ref{fig:Proliferation}a, the location of cells was visualized in both 2D and 3D. SEM images (Fig.~\ref{fig:Proliferation}b) of both scaffold's top and cross-section also further confirmed these information. Significant cell proliferation can be observed from day 7 to day 14 from both imaging methods. The results suggest that all scaffolds generated from the filtered dataset had the capacity to effectively support cell adhesion, proliferation along structures, and bridging between structures. In terms of analysis of the Alamar Blue assay, the fluorescence intensity is proportional to the amount of metabolically activated cells. As shown in Fig.~\ref{fig:Proliferation}c, the fluorescence intensity increased from day 1 to 14, suggesting that all these scaffolds are biocompatible without significant cytotoxicity. They also confirmed that all scaffolds' morphologies generated by our method were suitable for the usage of tissue engineering applications. Additionally, as can be observed, the fluorescence intensity increased with the volume fraction (VF). 
Notably, start from day 7, all optimized scaffolds presented higher values than conventional scaffolds, with VF=0.45 presented the significance, illustrated the advantage of geometry optimisation, particularly at the same VF. Within the optimized scaffold group, the fluorescence intensity increased with the VF, this could due to lower VF group induce higher amount of larger internal pores (above 600 um), which is less favoured by cell proliferation \cite{torres2017effect, murphy2010effect, o2005effect}. A similar trend was also observed for Saos-2 cells, as shown in Fig.~\ref{fig:Proliferation}d, ranging from day 1 to day 10, further confirmed our observation. The close readings among different designs on day 14 could due to the cell over-confluence at the later stage of proliferation. In summary, under the same volume fraction, optimized scaffolds presented significantly better cell proliferation results for both cell lines, confirming the significance to enhance the cell proliferation through scaffolds topological optimization, revealing the potential for future scaffold improvement.
Incorporating the results of compression tests already given in Fig.~\ref{fig:different_volume}, we can conclude that the scaffold structure generated by our method has been optimized for its mechanical and biological performance concurrently. 



%% file: discussion.tex
\section*{Discussion} 
This work presents a novel approach for generating microstructures with simultaneously optimized mechanical and biological performance, specifically targeting cell proliferation in bone tissue engineering. The challenge lies in the fact that optimizing biological performance cannot be achieved through numerical simulations, and traditional geometric metrics like surface area or surface-to-volume ratio are insufficient indicators of biological efficacy. To address this, we developed a data-driven morphology learning method that first learns the latent space of structures with superior biological performance and then applies FEA-based mechanical optimization within this space. Using this morphology learning-based approach, we can generate a scaffold design for bone tissue engineering that significantly improves both mechanical stiffness -- measured as compression modulus (by 29.69\%) and cell proliferation -- measured via Alamar Blue assay (by 37.05\% on Day 7 and 33.30\% on Day 14).

While the results from our experimental tests are promising, there is still room for improving the performance of our approach. In the current implementation, the dataset of structures with `good' biological performance was constructed by excluding those identified as having `poor' performance by a tissue engineering expert. This introduces potential bias stemming from the expert's existing knowledge. \revise{}{This may leave some structures with `poor' biological performance in the selected dataset, potentially reducing the effectiveness of our approach.} A more rigorous selection process would involve laboratory testing of all structures in the dataset\revise{, retaining only}{~to only retain} those \revise{that demonstrate}{with} superior biological performance \revise{}{in the selected dataset}. This will potentially lead to structures with better biological performance. Another limitation is that the sample structures are exclusively derived from TPMS although the shape variation has been enriched by interpolating between seven different TPMS types. We anticipate that incorporating a wider range of structure types (e.g., those inspired by nature \cite{SIDDIQUE2022natureInspire,BANDYOPADHYAY2021NatureInspire}) could further increase the diversity of structural shapes and enhance biological performance.

\revise{}{The effectiveness of our approach has been verified on a unit microstructure with specified dimensions. This microstructure can be periodically replicated and merged to occupy a larger region, which can then be trimmed into a specific bone shape. These operations can be efficiently performed using Boolean operations on the resultant structures generated by our approach, as they are represented as implicit solids defined by $f_\theta(\mathbf{x})$. The feasibility of additively manufacturing such periodically applied structures over large regions has been demonstrated in prior research\cite{ding2021stl}. Additionally, when fabricating the scaffold using FDM, water-soluble materials can be employed for support structures when necessary.}

From a broader perspective, this work introduces a general approach to integrating factors that cannot be directly evaluated through numerical simulations into structural TO, which has focused primarily on maximizing mechanical stiffness. By leveraging the data-driven morphology learning methodology as proposed in this paper, we can simultaneously optimize multiple objectives even when they are difficult to formulate mathematically. These objectives may include not only mechanical performance but also factors such as aesthetics, physiological comfort, psychological well-being, and biological compatibility. This approach opens up many new possibilities for designing structures that balance diverse and complex requirements, making it applicable to a broader range of fields from biomedical engineering to product design. For example, psychological and aesthetic aspects play a significant role in consumer product design, which can now be potentially addressed alongside structural optimization.

%% file: methods.tex
\section*{Methods}
\subsection*{Generation of structures using TPMS}
Seven types of TPMS used in the construction of dataset are defined by the following formulas~\cite{al2019multifunctional}: 
\begin{equation}\label{eq:tpms}
\begin{array}{l}
\phi_1 (\mathbf{x}) = \varphi (\mathbf{x})_P=\cos(X)+ \cos(Y)+\cos(Z),\\
\phi_2 (\mathbf{x}) = \varphi (\mathbf{x})_G=\sin(X)\cos(Y)+\sin(Y)\cos(Z)+\sin(Z)\cos(X),\\
\phi_3 (\mathbf{x}) = \varphi (\mathbf{x})_D=\cos(X)\cos(Y)\cos(Z)-\sin(X)\sin(Y)\sin(Z),\\
\phi_4 (\mathbf{x}) = \varphi (\mathbf{x})_{FKS}=\cos(2X)\sin(Y)\cos(Z)+\cos(X)\cos(2Y)\sin(Z)+\sin(X)\cos(Y)\cos(2Z),\\
\phi_5 (\mathbf{x}) = \varphi (\mathbf{x})_{IWP}=2[\cos(X)\cos(Y)+\cos(Y)\cos(Z)+\cos(Z)\cos(X)]-\cos(2X)-\cos(2Y)-\cos(2Z),\\
\phi_6 (\mathbf{x}) = \varphi (\mathbf{x})_{FRD}=4\cos(X)\cos(Y)\cos(Z)-[\cos(2X)\cos(2Y)+\cos(2Y)\cos(2Z)+\cos(2Z)\cos(2X)],\\
\phi_7 (\mathbf{x}) = \varphi (\mathbf{x})_{N}=3[\cos(X)+\cos(Y)+\cos(Z)]+4\cos(X)\cos(Y)\cos(Z),\\
\end{array}
\end{equation}
where $\mathbf{x} \in \mathbb{R}^3$ is the spatial coordinate of a point for each TPMS as an implicit function with $X=\frac{2\pi}{L} \mathbf{x}_x$, $Y=\frac{2\pi}{L} \mathbf{x}_y$ and $Z=\frac{2\pi}{L} \mathbf{x}_z$. Here $L$ is a coefficient to control the period of TPMS and $L=1.0$  is employed in our implementation to contain one complete period inside a unit cubic space. The sample models employed in our dataset are generated by blending these seven TPMS as 
\begin{equation}
    \phi(\mathbf{x})= \sum_{i=1}^7 w_i \phi_i (\mathbf{x}),
\end{equation}
where $w_{i, i=1,2,\ldots,7}$ are weighting factors, satisfying $\sum_{i} w_i \equiv 1$. Two different types of TPMS models are employed in our dataset: solids and sheets, where the boundary surface of a solid model is defined as $g(\mathbf{x})=\phi(\mathbf{x})-C_d=0$ with $C_d$ controlling the offsetting and the boundary surface of a sheet model is $g(\mathbf{x})= |\phi(\mathbf{x})| - C_t= 0$ with $C_t>0$ controlling the thickness. $g(\mathbf{x})<0$ and $g(\mathbf{x})>0$ denote the regions inside and outside a model respectively. When generating the dataset of morphology learning, Different offsetting values as $C_d=0.1, 0.2, \cdots, 1.0$ and $C_t=0.1, 0.2, \cdots, 0.9$ are employed for generating diverse models. Geometric evaluation of TPMS models is based on existing work in literature~\cite{ding2021stl,xu2024TPMS}. 

\subsection*{Neural network-based morphology learning}
\paragraph{Geometric representation}
During morphology learning, we represent each sample model in the dataset as a Signed Distance Function (SDF). For a given shape $H$ with its boundary surface described as $g(\mathbf{x})=0$, its SDF is defined as 
\begin{equation}\label{eq:sdf}
    s_H(\mathbf{x}) = \mathrm{sign}(g(\mathbf{x})) \min_{\mathbf{y}} \| \mathbf{x} -\mathbf{y} \| \quad s.t. \; g(\mathbf{y})=0,
\end{equation}
where $\mathbf{x},\mathbf{y} \in \mathbb{R}^3$ are 3D points and the constraint $g(\mathbf{y})=0$ defines the search range of $\mathbf{y}$ as among the boundary surface points of $H$. The SDF defined in Eq.~(\ref{eq:sdf}) is a query-based discrete function and typically non-differentiable. The morphology learning in our work trained a neural network to approximate SDFs of all models, conditioned on the latent code.

\begin{figure}[!t]
\centering
\includegraphics[width=0.75\linewidth]{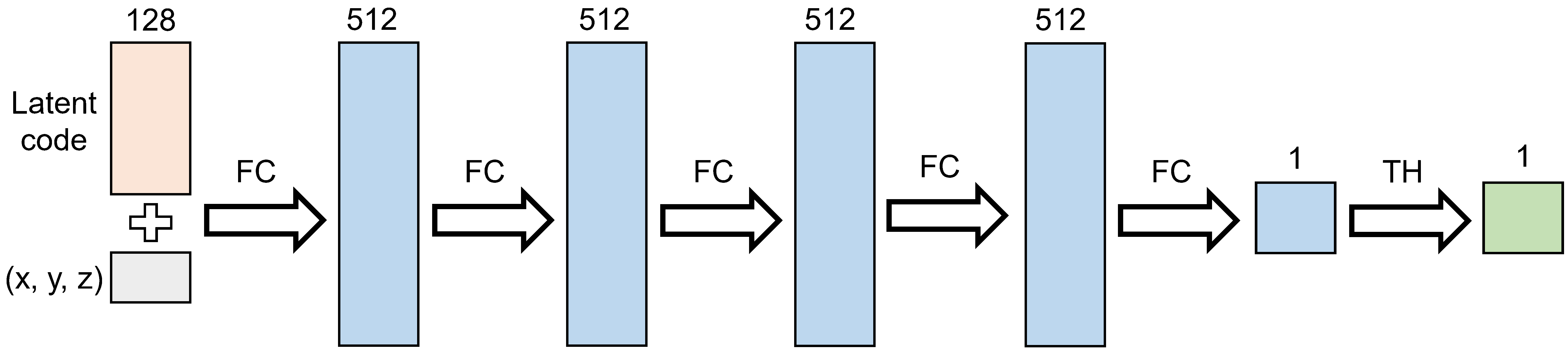}
\caption{\textbf{Network architecture of MLP used in morphology learning.}
}\label{fig:networkArch}
\end{figure}

\paragraph{Network architecture}
We use a Multi-Layer Perceptron (MLP) neural network for morphology learning, denoted by $f_\theta(\mathbf{x},\mathbf{z})$, where $\theta$ represents the network coefficients to be learned. Our implementation consists of 5 fully connected layers (FC), each with 512 neurons employing ReLU activation function, as illustrated in Fig.~\ref{fig:networkArch}. A final $\tanh(\cdot)$ function (TH) is applied at the network's output to project the results back into the interval $[-1.0,1.0]$. Unlike conventional SDFs, this neural SDF representation is differentiable, which enables the structural optimization conducted within the parameterized shape space. Further details will be provided in the following sections.

\paragraph{Loss function and training} 
To complete the training process of morphology learning, we first generate a set of random points in the spatial domain of a unit cube as $[0,1] \times [0,1] \times [0,1]$. The points are stored in a set $\mathcal{P}$, and $1,000,000$ points are employed in our implementation. Given a model $H$ of the filtered dataset $\mathcal{D}$, we evaluate the signed distance from every point $\mathbf{x} \in \mathcal{P}$ to $H$ as $s_H(\mathbf{x})$ by the method~\cite{baerentzen2005signed} presented in Eq.(\ref{eq:sdf}). With the help of these sample points, the loss function of training is defined as follows to learn an approximated SDF function $f_\theta$~\cite{park2019deepsdf}.
\begin{equation}
    L_{ML}(\{\mathbf{z}_{H}\}_{H\in\mathcal{D}}, \theta)=\sum_{H\in\mathcal{D}} \frac{1}{\#(\mathcal{P})} \sum_{\mathbf{x}\in\mathcal{P}}|f_{\theta}(\mathbf{x}, \mathbf{z}_{S})-s_H(\mathbf{x}))|+\lambda\sum_{H\in\mathcal{D}} \|\mathbf{z}_{H} \|^2,
\label{eq:loss}
\end{equation}
where the first term is a reconstruction term that minimizes the difference between the approximated and the actual signed distance values, and the second term is a regularization term that controls the magnitude of the $d$-dimensional latent code vector $\mathbf{z}_{H} \in \mathbb{R}^d$. $d=128$ is used in our implementation. $\lambda$ is a parameter used to balance between two terms in the loss function with $\lambda=10^{-4}$ chosen by experiment. The coefficients $\theta$ of the MLP neural network are unknown variables determined to encode the common shape patterns of the models in the dataset $\mathcal{D}$. 

\subsection*{Structure optimization in parameterized shape space}
Our approach of structural optimization, which preserves the shape patterns extracted from morphology learning, builds upon the density-based TO method SIMP. We first provide a brief introduction to SIMP. Key components of the SIMP framework will later be adapted for our pattern-preserving TO.

\paragraph{SIMP TO framework}\label{sec:simp} The optimization of mechanical stiffness is always formulated as a compliance minimization problem in the SIMP framework. The given design domain is discretized into finite elements and each element $e$ is associated a scalar value $\rho_e$ indicating the material of this element. The optimization result is expected to give either 0 (void) or 1 (solid) for the value of $\rho_e$. Then, the compliance minimization problem based on the SIMP method is formulated as:
\begin{eqnarray}
    \label{eq:top}
            & \min_{\boldsymbol{\rho}} &\mathbf{U}(\boldsymbol{\rho})^{\T}\mathbf{K}(\boldsymbol{\rho})\mathbf{U}(\boldsymbol{\rho})=\sum_{e=1}^{N} E_e(\rho_e)\mathbf{u}_e^{\T}\mathbf{k}_0\mathbf{u}_e,  \\
         &s.t. & \sum_{e=1}^{N} \rho_e v_e \leq V_0, \label{eq:Vol} \\
         &&\mathbf{K}(\boldsymbol{\rho})\mathbf{U}(\boldsymbol{\rho})=\mathbf{F}, \label{eq:FEA_in_SIMP}\\
         &&0\leq\rho_e\leq1, e=1,2,...,N,
\end{eqnarray}
where $N$ is the number of elements, $\mathbf{K(\boldsymbol{\rho})}$ and $\mathbf{U}(\boldsymbol{\rho})$ are the global stiffness matrix and nodal displacement vector, $V_0$ is the maximal available volume on the result, $v_e$ and $\mathbf{u}_e$ are the volume and nodal displacement of the element $e$, and $\mathbf{k}_0$ is the elemental stiffness matrix which is a constant. A common choice of FEA conducted for Eq.~(\ref{eq:FEA_in_SIMP}) in the SIMP framework is voxel-based FEA~\cite{liu2014efficient}. $E_e(.)$ is the interpolated Young’s modulus of element $e$ defined as
\begin{equation}
E_e(\rho_e) = E_{\min} + \rho_e^{p}(E_{\max}-E_{\min}),
\label{eq:SIMP}
\end{equation}
where $E_{\max}$ is the Young's modulus of the solid element, $E_{\min}$ is a small value to prevent the singularity of the global matrix $\mathbf{K}$, and $p$ is a penalization factor which is set as 3 in most cases. When updating the density values $\{ \rho_e\}$, a density filter~\cite{bendsoe2013topology} is also commonly applied to avoid the generation of checkerboard pattern as a mesh dependence problem~\cite{SigmundEtAl1998NumericalInstabilities}. The filtered density $\tilde{\rho}$ is obtained by
\begin{equation}
\tilde{\rho}_e = \frac{\sum_{i \in \mathcal{S}_e} w(\mathbf{x}_i,r_d)v_i \rho_i}{\sum_{i \in \mathcal{S}_e} w(\mathbf{x}_i,r_d) v_i},
\label{eq:filtering}
\end{equation}
with the weighting function $w(\mathbf{x}_i,r)$ being given as
\begin{equation}
\label{eq:linear-interpo}
w(\mathbf{x}_i,r) = r - ||\mathbf{x}_i - \mathbf{x}_e||.
\end{equation}
$r$ is the filter radius, and the set $\mathcal{S}_e$ defines the neighboring elements of $e$ as $\mathcal{S}_e = \{i ~|~ w(\mathbf{x}_i,r) > 0\}$. $\mathbf{x}_e$ and $\mathbf{x}_i$ are the positions of the centroid of elements $e$ and $i$. In the SIMP framework, the problem is solved by iteratively updating the density-values according to the sensitivity analysis and the density filter, employing either the Optimality Criteria (OC) scheme~\cite{bendsoe1995optimization} or the Method of Moving Asymptotes (MMA) scheme~\cite{svanberg1987method}.

\paragraph{Pattern-preserving TO framework}\label{sec:tof}
Different from density-based TO framework, the design variables of our pattern-preserving TO are the latent codes $\mathbf{z}$ of a network $f_\theta(\mathbf{x},\mathbf{z})$. The design domain is again discretized into voxel elements. Specifically, we wish to assign the density $\rho_e = 1.0$ for an element $e$ if its center $\mathbf{x}_e$ is located inside the solid represented by a latent code $\mathbf{z}$, which can be detected by $\tilde{s}_e= f_\theta(\mathbf{x},\mathbf{z}) \leq 0$. When $\tilde{s}_e >0$, the element $e$ is considered as a void assigned with $\rho_e = 0.0$. This is realized by introducing following two projections. 

\begin{figure}[t]
\centering
\includegraphics[width=1\linewidth]{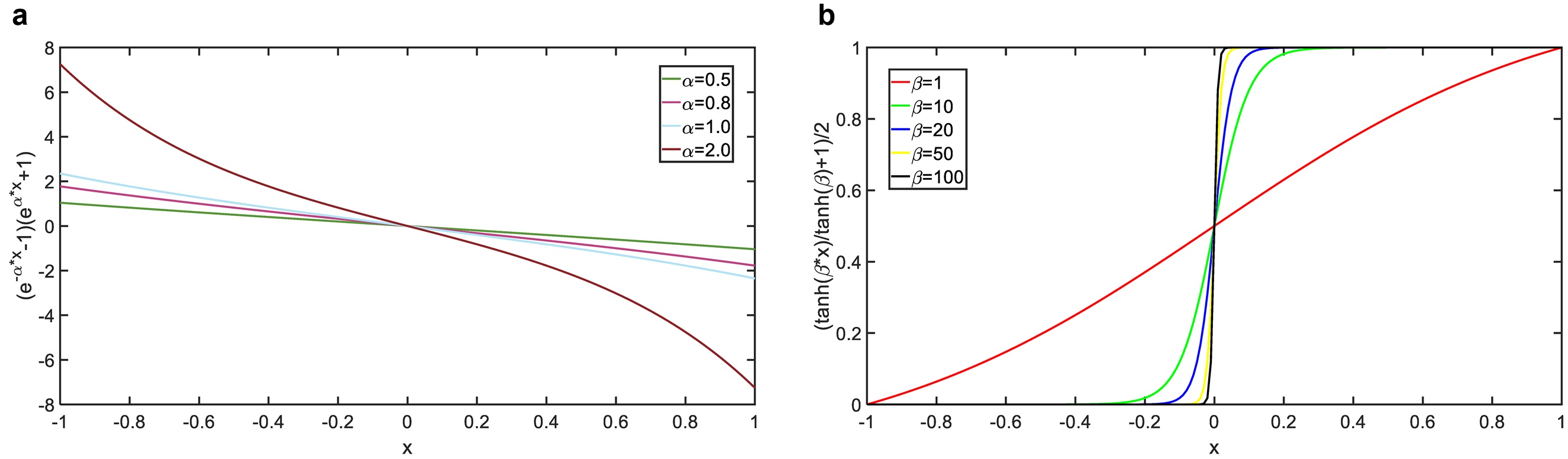}
\caption{\textbf{The illustration of projection functions used in our pattern-preserving TO.} 
\textbf{a} The function as defined in Eq.(\ref{eq:proj1}). 
\textbf{b} The projection as defined in Eq.(\ref{eq:proj2}).
}\label{fig:projection}
\end{figure}

First of all, in order to avoid the results with many `gray' elements with densities between 0 and 1 (especially those nearly around \textit{zero}), we project the value of $s_e$ to shift it away from zero by
\begin{equation}\label{eq:proj1}
   \bar{\rho}_e=\left(e^{\alpha (-\tilde{s}_e)} - 1\right)\left(e^{\alpha (\tilde{s}_e)} + 1\right), 
\end{equation}
where $\alpha$ is a parameter to control the sharpness of the projection. The sharpness of this projection function in Eq.~(\ref{eq:proj1}) can be controlled by using different values of $\alpha$ as shown in Fig.~\ref{fig:projection}a. $\alpha=2.0$ is chosen in our implementation according to experiments. Then, we further shift the values onto 0 or 1 by 
\begin{equation}\label{eq:proj2}
   \rho_e=\frac{1}{2}\left(\frac{\tanh(\beta \bar{\rho}_e)}{\tanh(\beta)}+ 1\right),
\end{equation}
with $\beta$ being a parameter to control the sharpness of the projection (see Fig.~\ref{fig:projection}b for the shapes of different $\beta$s). As we can see, increasing the value of $\beta$ will give sharper corresponding curves. To achieve a good convergence in computation, we start from $\beta=1.0$ and increase its value by $\beta=4.0$ after every 30 iterations until it reaches 128.

With the help of these projection operators, we define the pattern-preserving TO as the following minimization problem:
\begin{eqnarray}
    \label{eq:framework}
          & \min_{\mathbf{\mathbf{z}}} &c(\mathbf{z})=\mathbf{U}(\boldsymbol{\rho}(\mathbf{z}))^T\mathbf{K}(\boldsymbol{\rho}(\mathbf{z}))\mathbf{U}(\boldsymbol{\rho}(\mathbf{z})),  \\
         &s.t. & V=\sum_{e=1}^{N} \rho_e(\mathbf{z}) v_e \leq V_0, \label{eq:NNTOVolConstraint} \\
         &&\mathbf{K(\boldsymbol{\rho}(\mathbf{z}))U(\boldsymbol{\rho}(\mathbf{z}))}=\mathbf{F}, \\
         &&\mathbf{\tilde{s}}=f_{\theta}(\mathbf{z}), \\
         &&\mathbf{\bar{\rho}}=e^{\alpha \mathbf{\tilde{s}}} - 1, \\
         &&\mathbf{\rho}=\frac{1}{2}\left(\frac{\tanh(\beta \mathbf{\bar{\rho}})}{\tanh(\beta)}+ 1\right).
\end{eqnarray}
Different from SIMP framework that usually takes either OC or MMA scheme to update the values of design variables, the design variables $\mathbf{z}$ in our approach are updated by a pipeline of neural network-based self-learning. Specifically, the Adam optimizer~\cite{kingma2014adam} is employed to determine the values of $\mathbf{z}$ via backpropagation as illustrated in Fig.\ref{fig:TOBackpropagation}. 

The following loss function combining the objective function $c(\mathbf{z})$ and the volume constraint (i.e., Eq.(\ref{eq:NNTOVolConstraint})) is defined to govern the self-learning (i.e., optimization) as 
\begin{equation}
   L_{TO}=\frac{c(\mathbf{z})}{c_0}+ \mu \left(\frac{V}{V_0}-1\right)^2,
   \label{eq:loss}
\end{equation}
where $c_0$ is the compliance of the initial structure, and $\mu$ is a parameter to balance the objective function and volume constraint. In our implementation, we start the computation with $\mu=10$ and increase its value by $1$ per iteration until it reaches to 500. The derivative of the loss function $L_{TO}$ regarding to the latent code vector $\mathbf{z}$ can be computed by using the chain rule as
\begin{equation}
    \label{eq:deri}
         \frac{\partial L_{TO}}{\partial{z_i}}=\sum_{e=1}^{N}\frac{\partial L_{TO}}{\partial \rho_e}\frac{\partial \rho_e}{\partial \bar{\rho}_e}\frac{\partial \bar{\rho}_e}{\partial \tilde{s}_e}\frac{\partial \tilde{s}_e}{\partial z_i}
\end{equation}
with $z_i$ being the $i-th$ component of the latent code vector $\mathbf{z}$. We have
\begin{equation}
\frac{\partial L_{TO}}{\partial \rho_e}=\frac{\partial c}{c_0\partial \rho_e}+2\frac{\mu}{V_0}\left(\frac{V}{V_0}-1\right)\frac{\partial V}{\partial \rho_e},
\end{equation}
where $\frac{\partial c}{\partial \rho_e}$ can be obtained by the adjoint analysis~\cite{bendsoe2013topology} and $\frac{\partial V}{\partial \rho_e}$ can be directly calculated using Eq.~(\ref{eq:NNTOVolConstraint}). Analytical form of $\frac{\partial \bar{\rho}_e}{\partial \tilde{s}_e}$ and $\frac{\partial \rho_e}{\partial \bar{\rho}_e}$ can be obtained from Eq.~(\ref{eq:proj1}) and Eq.~(\ref{eq:proj2}) respectively. Since the neural network $f_{\theta}(\cdot)$ is differentiable, $\frac{\partial \tilde{s}_e}{\partial z_i}$ can be calculated with $\tilde{s}_e=f_{\theta}(\mathbf{x}_e, \mathbf{z})$ through automatic differentiation. In conclusion, the loss function $L$ is differentiable, which allows to use the automatic differentiation function of Adam optimizer. 

The pipeline of our pattern-preserving TO framework has been given in Fig.\ref{fig:TOBackpropagation}. After determining the optimized latent code $\mathbf{z}^{opt}$, the SDF of the resultant structure $f_\theta(\mathbf{x},\mathbf{z}^{opt})$ is employed to generate the toolpath for 3D printing\cite{ding2021stl}. 

\subsection*{Filament deposition based 3D printing for scaffold fabrication}
Biocompatiable and biodegradable polymeric material -- Polycarpolacton (PCL) with average molecular weight as 40,000 Da 
and Kywoo Tycoon Max 3D Printer equipped with a 0.2 mm nozzle were employed in our experiments for scaffold fabrication. Cura software (Ultimaker, Netherlands) was used for slicing, key parameters considered in fabrication include layer thickness = 0.1 mm, printing temperature = 85$^\circ$C, and printing velocity = 30 mm/s. The scaffolds were printed in the dimension of 12.0 mm × 12.0 mm × 18.0 mm (W × D × H) for mechanical characterization and the dimension of 12.0 mm × 12.0 mm × 6.0 mm (W × D × H) for biological characterization. 

\subsection*{Mechanical characterization}
Uni-axial mechanical compression tests as shown in Fig.~\ref{fig:different_volume}d were conducted to evaluate the mechanical properties (compressive modulus) of all 3D printed scaffolds. Instron 3344 single-column table frame system with 100N load cell was employed for the compression test. 
All tests were performed in dry conditions with 0.5 mm/min compression rate and 0.2 mm/mm (20\%) strain limit. The resultant mechanical stress $\sigma$ and strain $\epsilon$ were used to calculate the compressive modulus following the methodology outlined by Fiedler \textit{et al} \cite{fiedler2015mechanical}, in the Origin software (OriginLab, USA).

\subsection*{Biological characterization}
The effectiveness of the optimized scaffold structures was evaluated through biological characterization and compared to conventional scaffolds. The specific methodologies of these comparisons are detailed as follows.

\paragraph{hADSC culture and seeding}
Human adipose-derived stem cells (hADSCs, passage 6-8, StemPro\textregistered, Invitrogen, USA) were subcultured in MesenPRO RS\texttrademark~basal medium (Invitrogen, USA) within T75 flasks (Sigma-Aldrich, UK) under controlled conditions. Cells were harvested at approximately 85\% confluence using 0.05\% trypsin-EDTA (Invitrogen, USA) for detachment. Prior to cell seeding, each scaffold underwent a sterilization protocol, initially immersed in 70\% ethanol followed by multiple washes with phosphate-buffered saline (PBS, Sigma-Aldrich, UK) to ensure complete removal of ethanol. The scaffolds were then air-dried under sterile conditions overnight. For the cell seeding process, approximately 50,000 hADSCs suspended in 1.2 mL of medium were introduced to each scaffold within 24-well plates. Scaffolds were cultured under standard incubator conditions (37$^\circ$C, 95\% humidity, and 5\% CO$_2$ concentration). The culture medium was refreshed every 2 days to maintain nutrient availability and promote cell proliferation and differentiation over the experimental period.

\paragraph{Cell proliferation analysis}
hADSCs proliferation was evaluated through Alamar Blue assay after 1, 3, 5, 7, 10, and 14 days of cell seeding, using resazurin sodium salt (Invitrogen, USA). At each time point, each cell-seeded scaffold was transferred to a new well and added with 1.2 mL medium containing 0.001\% resazurin sodium salt. After 4 hours of incubation under standard conditions in the dark, 150 $\upmu$L medium was transferred to a 96-well plate, and the fluorescence intensity was measured using a TECAN Infinite 200 plate reader (Tecan, Switzerland) at E\textsubscript{x}/E\textsubscript{m} = 540/590 nm.

\paragraph{Bioimaging}
Confocal microscopy imaging was conducted to evaluate cell adhesion and spreading status on the scaffolds. On day 7 and 14 of the cell culture, cell-seeded scaffolds were fixed with 10\% formalin (Sigma-Aldrich, UK), permeabilized with 0.1\% Triton X-100 (Sigma-Aldrich, UK) in PBS, and stained with Alexa Fluor\texttrademark~488 Phalloidin (E\textsubscript{x}/E\textsubscript{m} = 490/525 nm, Invitrogen, USA) and 4',6-Diamidino-2-phenylindole dihydrochloride (E\textsubscript{x}/E\textsubscript{m} = 350/470 nm, DAPI; Invitrogen, USA) according to manufacturer's guidence. 
Confocal images were captured using a Leica SP8 LIGHTNING confocal microscope (Leica, Germany) with 405 nm and 499 nm lasers.
Scanning electron microscopy (SEM) was used to further evaluate cell attachment and proliferation at days 7 and 14. The fixed cell-seeded scaffolds were dehydrated by gradient ethanol (50\%, 70\%, 80\%, 90\%, and 100\%), followed by the treatment of 1:1 ethanol and hexamethyldisilazane (HMDS, Sigma-Aldrich, UK) and pure HMDS, considering 15 mins each step. The dehydrated samples were air-dried for volatilization, surface coated with 6~nm gold-palladium (4:1), and imaged using a TESCAN MIRA3 SEM (TESCAN, Czech) at an accelerating voltage of 2~kV. 

\subsection*{Data analysis} 
All mechanical and biological experiments were conducted at least three times and the results were reported as mean value with standard deviation. Origin software (OriginLab, USA) was used for data analysis, and one-way analysis of variance (one-way ANOVA) with Tukey \textit{post hoc} test was applied for statistical analyses by setting significance levels as * p < 0.05, ** p < 0.01, and *** p < 0.001. 

%% file: main.bbl
\begin{thebibliography}{10}
\urlstyle{rm}
\expandafter\ifx\csname url\endcsname\relax
  \def\url#1{\texttt{#1}}\fi
\expandafter\ifx\csname urlprefix\endcsname\relax\def\urlprefix{URL }\fi
\expandafter\ifx\csname doiprefix\endcsname\relax\def\doiprefix{DOI: }\fi
\providecommand{\bibinfo}[2]{#2}
\providecommand{\eprint}[2][]{\url{#2}}

\bibitem{hollister2005porous}
\bibinfo{author}{Hollister, S.~J.}
\newblock \bibinfo{journal}{\bibinfo{title}{Porous scaffold design for tissue engineering}}.
\newblock {\emph{\JournalTitle{Nature materials}}} \textbf{\bibinfo{volume}{4}}, \bibinfo{pages}{518--524}, \doiprefix\url{https://doi.org/10.1038/nmat1421} (\bibinfo{year}{2005}).

\bibitem{chen2011microstructure}
\bibinfo{author}{Chen, Y.}, \bibinfo{author}{Zhou, S.} \& \bibinfo{author}{Li, Q.}
\newblock \bibinfo{journal}{\bibinfo{title}{Microstructure design of biodegradable scaffold and its effect on tissue regeneration}}.
\newblock {\emph{\JournalTitle{Biomaterials}}} \textbf{\bibinfo{volume}{32}}, \bibinfo{pages}{5003--5014}, \doiprefix\url{https://doi.org/10.1016/j.biomaterials.2011.03.064} (\bibinfo{year}{2011}).

\bibitem{zhang2019three}
\bibinfo{author}{Zhang, L.}, \bibinfo{author}{Yang, G.}, \bibinfo{author}{Johnson, B.~N.} \& \bibinfo{author}{Jia, X.}
\newblock \bibinfo{journal}{\bibinfo{title}{Three-dimensional (3d) printed scaffold and material selection for bone repair}}.
\newblock {\emph{\JournalTitle{Acta biomaterialia}}} \textbf{\bibinfo{volume}{84}}, \bibinfo{pages}{16--33}, \doiprefix\url{https://doi.org/10.1016/j.actbio.2018.11.039} (\bibinfo{year}{2019}).

\bibitem{metz2020towards}
\bibinfo{author}{Metz, C.}, \bibinfo{author}{Duda, G.~N.} \& \bibinfo{author}{Checa, S.}
\newblock \bibinfo{journal}{\bibinfo{title}{Towards multi-dynamic mechano-biological optimization of 3d-printed scaffolds to foster bone regeneration}}.
\newblock {\emph{\JournalTitle{Acta biomaterialia}}} \textbf{\bibinfo{volume}{101}}, \bibinfo{pages}{117--127}, \doiprefix\url{https://doi.org/10.1016/j.actbio.2019.10.029} (\bibinfo{year}{2020}).

\bibitem{kumar2016low}
\bibinfo{author}{Kumar, A.} \emph{et~al.}
\newblock \bibinfo{journal}{\bibinfo{title}{Low temperature additive manufacturing of three dimensional scaffolds for bone-tissue engineering applications: Processing related challenges and property assessment}}.
\newblock {\emph{\JournalTitle{Materials Science and Engineering: R: Reports}}} \textbf{\bibinfo{volume}{103}}, \bibinfo{pages}{1--39}, \doiprefix\url{https://doi.org/10.1016/j.mser.2016.01.001} (\bibinfo{year}{2016}).

\bibitem{wang2016topological}
\bibinfo{author}{Wang, X.} \emph{et~al.}
\newblock \bibinfo{journal}{\bibinfo{title}{Topological design and additive manufacturing of porous metals for bone scaffolds and orthopaedic implants: A review}}.
\newblock {\emph{\JournalTitle{Biomaterials}}} \textbf{\bibinfo{volume}{83}}, \bibinfo{pages}{127--141}, \doiprefix\url{https://doi.org/10.1016/j.biomaterials.2016.01.012} (\bibinfo{year}{2016}).

\bibitem{charbonnier2021additive}
\bibinfo{author}{Charbonnier, B.}, \bibinfo{author}{Hadida, M.} \& \bibinfo{author}{Marchat, D.}
\newblock \bibinfo{journal}{\bibinfo{title}{Additive manufacturing pertaining to bone: Hopes, reality and future challenges for clinical applications}}.
\newblock {\emph{\JournalTitle{Acta Biomaterialia}}} \textbf{\bibinfo{volume}{121}}, \bibinfo{pages}{1--28}, \doiprefix\url{https://doi.org/10.1016/j.actbio.2020.11.039} (\bibinfo{year}{2021}).

\bibitem{moroni20063d}
\bibinfo{author}{Moroni, L.}, \bibinfo{author}{De~Wijn, J.} \& \bibinfo{author}{Van~Blitterswijk, C.}
\newblock \bibinfo{journal}{\bibinfo{title}{3d fiber-deposited scaffolds for tissue engineering: influence of pores geometry and architecture on dynamic mechanical properties}}.
\newblock {\emph{\JournalTitle{Biomaterials}}} \textbf{\bibinfo{volume}{27}}, \bibinfo{pages}{974--985}, \doiprefix\url{https://doi.org/10.1016/j.biomaterials.2005.07.023} (\bibinfo{year}{2006}).

\bibitem{zhang2020biomechanical}
\bibinfo{author}{Zhang, X.-Y.}, \bibinfo{author}{Yan, X.-C.}, \bibinfo{author}{Fang, G.} \& \bibinfo{author}{Liu, M.}
\newblock \bibinfo{journal}{\bibinfo{title}{Biomechanical influence of structural variation strategies on functionally graded scaffolds constructed with triply periodic minimal surface}}.
\newblock {\emph{\JournalTitle{Additive Manufacturing}}} \textbf{\bibinfo{volume}{32}}, \bibinfo{pages}{101015}, \doiprefix\url{https://doi.org/10.1016/j.addma.2019.101015} (\bibinfo{year}{2020}).

\bibitem{foroughi2022multi}
\bibinfo{author}{Foroughi, A.~H.} \& \bibinfo{author}{Razavi, M.~J.}
\newblock \bibinfo{journal}{\bibinfo{title}{Multi-objective shape optimization of bone scaffolds: Enhancement of mechanical properties and permeability}}.
\newblock {\emph{\JournalTitle{Acta Biomaterialia}}} \textbf{\bibinfo{volume}{146}}, \bibinfo{pages}{317--340}, \doiprefix\url{https://doi.org/10.1016/j.actbio.2022.04.051} (\bibinfo{year}{2022}).

\bibitem{van20223d}
\bibinfo{author}{Van~Hede, D.} \emph{et~al.}
\newblock \bibinfo{journal}{\bibinfo{title}{3d-printed synthetic hydroxyapatite scaffold with in silico optimized macrostructure enhances bone formation in vivo}}.
\newblock {\emph{\JournalTitle{Advanced Functional Materials}}} \textbf{\bibinfo{volume}{32}}, \bibinfo{pages}{2105002}, \doiprefix\url{https://doi.org/10.1002/adfm.202105002} (\bibinfo{year}{2022}).

\bibitem{jin2024precision}
\bibinfo{author}{Jin, J.} \emph{et~al.}
\newblock \bibinfo{journal}{\bibinfo{title}{Precision pore structure optimization of additive manufacturing porous tantalum scaffolds for bone regeneration: A proof-of-concept study}}.
\newblock {\emph{\JournalTitle{Biomaterials}}} \bibinfo{pages}{122756}, \doiprefix\url{https://doi.org/10.1016/j.biomaterials.2024.122756} (\bibinfo{year}{2024}).

\bibitem{park2019deepsdf}
\bibinfo{author}{Park, J.~J.}, \bibinfo{author}{Florence, P.}, \bibinfo{author}{Straub, J.}, \bibinfo{author}{Newcombe, R.} \& \bibinfo{author}{Lovegrove, S.}
\newblock \bibinfo{title}{{DeepSDF}: Learning continuous signed distance functions for shape representation}.
\newblock In \emph{\bibinfo{booktitle}{Proceedings of the IEEE/CVF conference on computer vision and pattern recognition}}, \bibinfo{pages}{165--174}, \doiprefix\url{https://doi.org/10.1109/CVPR.2019.00025} (\bibinfo{year}{2019}).

\bibitem{hui2022neural}
\bibinfo{author}{Hui, K.-H.}, \bibinfo{author}{Li, R.}, \bibinfo{author}{Hu, J.} \& \bibinfo{author}{Fu, C.-W.}
\newblock \bibinfo{title}{Neural template: Topology-aware reconstruction and disentangled generation of 3d meshes}.
\newblock In \emph{\bibinfo{booktitle}{Proceedings of the IEEE/CVF conference on computer vision and pattern recognition}}, \bibinfo{pages}{18572--18582}, \doiprefix\url{https://doi.org/10.1109/CVPR52688.2022.01802} (\bibinfo{year}{2022}).

\bibitem{abbas2023deepmorpher}
\bibinfo{author}{Abbas, A.}, \bibinfo{author}{Rafiee, A.} \& \bibinfo{author}{Haase, M.}
\newblock \bibinfo{journal}{\bibinfo{title}{{DeepMorpher}: deep learning-based design space dimensionality reduction for shape optimisation}}.
\newblock {\emph{\JournalTitle{Journal of Engineering Design}}} \textbf{\bibinfo{volume}{34}}, \bibinfo{pages}{254--270}, \doiprefix\url{https://doi.org/10.1080/09544828.2023.2192606} (\bibinfo{year}{2023}).

\bibitem{Guillard2024}
\bibinfo{author}{Guillard, B.} \emph{et~al.}
\newblock \bibinfo{journal}{\bibinfo{title}{Deepmesh: Differentiable iso-surface extraction}}.
\newblock {\emph{\JournalTitle{IEEE Transactions on Pattern Analysis and Machine Intelligence}}} \bibinfo{pages}{1--15}, \doiprefix\url{https://doi.org/10.1109/TPAMI.2024.3392291} (\bibinfo{year}{2024}).

\bibitem{siddique2022lessons}
\bibinfo{author}{Siddique, S.~H.}, \bibinfo{author}{Hazell, P.~J.}, \bibinfo{author}{Wang, H.}, \bibinfo{author}{Escobedo, J.~P.} \& \bibinfo{author}{Ameri, A.~A.}
\newblock \bibinfo{journal}{\bibinfo{title}{Lessons from nature: {3D} printed bio-inspired porous structures for impact energy absorption--a review}}.
\newblock {\emph{\JournalTitle{Additive Manufacturing}}} \textbf{\bibinfo{volume}{58}}, \bibinfo{pages}{103051}, \doiprefix\url{https://doi.org/0.1016/j.addma.2022.103051} (\bibinfo{year}{2022}).

\bibitem{bendsoe2013topology}
\bibinfo{author}{Bends{\o}e, M.~P.} \& \bibinfo{author}{Sigmund, O.}
\newblock \emph{\bibinfo{title}{Topology optimization: theory, methods, and applications}} (\bibinfo{publisher}{Springer Science \& Business Media}, \bibinfo{year}{2013}).

\bibitem{aage2017giga}
\bibinfo{author}{Aage, N.}, \bibinfo{author}{Andreassen, E.}, \bibinfo{author}{Lazarov, B.~S.} \& \bibinfo{author}{Sigmund, O.}
\newblock \bibinfo{journal}{\bibinfo{title}{Giga-voxel computational morphogenesis for structural design}}.
\newblock {\emph{\JournalTitle{Nature}}} \textbf{\bibinfo{volume}{550}}, \bibinfo{pages}{84--86}, \doiprefix\url{https://doi.org/10.1038/nature23911} (\bibinfo{year}{2017}).

\bibitem{allaire2004structural}
\bibinfo{author}{Allaire, G.}, \bibinfo{author}{Jouve, F.} \& \bibinfo{author}{Toader, A.-M.}
\newblock \bibinfo{journal}{\bibinfo{title}{Structural optimization using sensitivity analysis and a level-set method}}.
\newblock {\emph{\JournalTitle{Journal of computational physics}}} \textbf{\bibinfo{volume}{194}}, \bibinfo{pages}{363--393}, \doiprefix\url{https://doi.org/10.1016/j.jcp.2003.09.032} (\bibinfo{year}{2004}).

\bibitem{wang2003level}
\bibinfo{author}{Wang, M.~Y.}, \bibinfo{author}{Wang, X.} \& \bibinfo{author}{Guo, D.}
\newblock \bibinfo{journal}{\bibinfo{title}{A level set method for structural topology optimization}}.
\newblock {\emph{\JournalTitle{Computer methods in applied mechanics and engineering}}} \textbf{\bibinfo{volume}{192}}, \bibinfo{pages}{227--246}, \doiprefix\url{https://doi.org/10.1016/S0045-7825(02)00559-5} (\bibinfo{year}{2003}).

\bibitem{borrvall2003topology}
\bibinfo{author}{Borrvall, T.} \& \bibinfo{author}{Petersson, J.}
\newblock \bibinfo{journal}{\bibinfo{title}{Topology optimization of fluids in stokes flow}}.
\newblock {\emph{\JournalTitle{International journal for numerical methods in fluids}}} \textbf{\bibinfo{volume}{41}}, \bibinfo{pages}{77--107}, \doiprefix\url{https://doi.org/10.1002/fld.426} (\bibinfo{year}{2003}).

\bibitem{zhou2008variational}
\bibinfo{author}{Zhou, S.} \& \bibinfo{author}{Li, Q.}
\newblock \bibinfo{journal}{\bibinfo{title}{A variational level set method for the topology optimization of steady-state navier--stokes flow}}.
\newblock {\emph{\JournalTitle{Journal of Computational Physics}}} \textbf{\bibinfo{volume}{227}}, \bibinfo{pages}{10178--10195}, \doiprefix\url{https://doi.org/10.1016/j.jcp.2008.08.022} (\bibinfo{year}{2008}).

\bibitem{ha2005topological}
\bibinfo{author}{Ha, S.-H.} \& \bibinfo{author}{Cho, S.}
\newblock \bibinfo{journal}{\bibinfo{title}{Topological shape optimization of heat conduction problems using level set approach}}.
\newblock {\emph{\JournalTitle{Numerical Heat Transfer, Part B: Fundamentals}}} \textbf{\bibinfo{volume}{48}}, \bibinfo{pages}{67--88}, \doiprefix\url{https://doi.org/10.1080/10407790590935966} (\bibinfo{year}{2005}).

\bibitem{yamada2011level}
\bibinfo{author}{Yamada, T.}, \bibinfo{author}{Izui, K.} \& \bibinfo{author}{Nishiwaki, S.}
\newblock \bibinfo{journal}{\bibinfo{title}{A level set-based topology optimization method for maximizing thermal diffusivity in problems including design-dependent effects}}.
\newblock {\emph{\JournalTitle{Journal of Mechanical Design}}} \textbf{\bibinfo{volume}{133}}, \bibinfo{pages}{031011}, \doiprefix\url{https://doi.org/10.1115/1.4003684} (\bibinfo{year}{2011}).

\bibitem{zhuang2007level}
\bibinfo{author}{Zhuang, C.}, \bibinfo{author}{Xiong, Z.} \& \bibinfo{author}{Ding, H.}
\newblock \bibinfo{journal}{\bibinfo{title}{A level set method for topology optimization of heat conduction problem under multiple load cases}}.
\newblock {\emph{\JournalTitle{Computer methods in applied mechanics and engineering}}} \textbf{\bibinfo{volume}{196}}, \bibinfo{pages}{1074--1084}, \doiprefix\url{https://doi.org/10.1016/j.cma.2006.08.005} (\bibinfo{year}{2007}).

\bibitem{zhu2016topology}
\bibinfo{author}{Zhu, J.-H.}, \bibinfo{author}{Zhang, W.-H.} \& \bibinfo{author}{Xia, L.}
\newblock \bibinfo{journal}{\bibinfo{title}{Topology optimization in aircraft and aerospace structures design}}.
\newblock {\emph{\JournalTitle{Archives of computational methods in engineering}}} \textbf{\bibinfo{volume}{23}}, \bibinfo{pages}{595--622}, \doiprefix\url{https://doi.org/10.1007/s11831-015-9151-2} (\bibinfo{year}{2016}).

\bibitem{guanghui2020aerospace}
\bibinfo{author}{Guanghui, S.} \emph{et~al.}
\newblock \bibinfo{journal}{\bibinfo{title}{An aerospace bracket designed by thermo-elastic topology optimization and manufactured by additive manufacturing}}.
\newblock {\emph{\JournalTitle{Chinese Journal of Aeronautics}}} \textbf{\bibinfo{volume}{33}}, \bibinfo{pages}{1252--1259}, \doiprefix\url{https://doi.org/10.1016/j.cja.2019.09.006} (\bibinfo{year}{2020}).

\bibitem{jankovics2019customization}
\bibinfo{author}{Jankovics, D.} \& \bibinfo{author}{Barari, A.}
\newblock \bibinfo{journal}{\bibinfo{title}{Customization of automotive structural components using additive manufacturing and topology optimization}}.
\newblock {\emph{\JournalTitle{IFAC-PapersOnLine}}} \textbf{\bibinfo{volume}{52}}, \bibinfo{pages}{212--217}, \doiprefix\url{https://doi.org/10.1016/j.ifacol.2019.10.066} (\bibinfo{year}{2019}).

\bibitem{jewett2019topology}
\bibinfo{author}{Jewett, J.~L.} \& \bibinfo{author}{Carstensen, J.~V.}
\newblock \bibinfo{journal}{\bibinfo{title}{Topology-optimized design, construction and experimental evaluation of concrete beams}}.
\newblock {\emph{\JournalTitle{Automation in Construction}}} \textbf{\bibinfo{volume}{102}}, \bibinfo{pages}{59--67}, \doiprefix\url{https://doi.org/10.1016/j.autcon.2019.02.001} (\bibinfo{year}{2019}).

\bibitem{xie1993simple}
\bibinfo{author}{Xie, Y.~M.} \& \bibinfo{author}{Steven, G.~P.}
\newblock \bibinfo{journal}{\bibinfo{title}{A simple evolutionary procedure for structural optimization}}.
\newblock {\emph{\JournalTitle{Computers \& structures}}} \textbf{\bibinfo{volume}{49}}, \bibinfo{pages}{885--896}, \doiprefix\url{https://doi.org/10.1016/0045-7949(93)90035-C} (\bibinfo{year}{1993}).

\bibitem{guo2014doing}
\bibinfo{author}{Guo, X.}, \bibinfo{author}{Zhang, W.} \& \bibinfo{author}{Zhong, W.}
\newblock \bibinfo{journal}{\bibinfo{title}{Doing topology optimization explicitly and geometrically—a new moving morphable components based framework}}.
\newblock {\emph{\JournalTitle{Journal of Applied Mechanics}}} \textbf{\bibinfo{volume}{81}}, \bibinfo{pages}{081009}, \doiprefix\url{https://doi.org/10.1115/1.4027609} (\bibinfo{year}{2014}).

\bibitem{garner2019compatibility}
\bibinfo{author}{Garner, E.}, \bibinfo{author}{Kolken, H.~M.}, \bibinfo{author}{Wang, C.~C.}, \bibinfo{author}{Zadpoor, A.~A.} \& \bibinfo{author}{Wu, J.}
\newblock \bibinfo{journal}{\bibinfo{title}{Compatibility in microstructural optimization for additive manufacturing}}.
\newblock {\emph{\JournalTitle{Additive Manufacturing}}} \textbf{\bibinfo{volume}{26}}, \bibinfo{pages}{65--75}, \doiprefix\url{https://doi.org/10.1016/j.addma.2018.12.007} (\bibinfo{year}{2019}).

\bibitem{liu2024ultrastiff}
\bibinfo{author}{Liu, Y.} \emph{et~al.}
\newblock \bibinfo{journal}{\bibinfo{title}{Ultrastiff metamaterials generated through a multilayer strategy and topology optimization}}.
\newblock {\emph{\JournalTitle{Nature Communications}}} \textbf{\bibinfo{volume}{15}}, \bibinfo{pages}{2984}, \doiprefix\url{https://doi.org/10.1038/s41467-024-47089-8} (\bibinfo{year}{2024}).

\bibitem{woldseth2022use}
\bibinfo{author}{Woldseth, R.~V.}, \bibinfo{author}{Aage, N.}, \bibinfo{author}{B{\ae}rentzen, J.~A.} \& \bibinfo{author}{Sigmund, O.}
\newblock \bibinfo{journal}{\bibinfo{title}{On the use of artificial neural networks in topology optimisation}}.
\newblock {\emph{\JournalTitle{Structural and Multidisciplinary Optimization}}} \textbf{\bibinfo{volume}{65}}, \bibinfo{pages}{294}, \doiprefix\url{https://doi.org/10.1007/s00158-022-03347-1} (\bibinfo{year}{2022}).

\bibitem{abueidda2020topology}
\bibinfo{author}{Abueidda, D.~W.}, \bibinfo{author}{Koric, S.} \& \bibinfo{author}{Sobh, N.~A.}
\newblock \bibinfo{journal}{\bibinfo{title}{Topology optimization of 2d structures with nonlinearities using deep learning}}.
\newblock {\emph{\JournalTitle{Computers \& Structures}}} \textbf{\bibinfo{volume}{237}}, \bibinfo{pages}{106283}, \doiprefix\url{https://doi.org/10.1016/j.compstruc.2020.106283} (\bibinfo{year}{2020}).

\bibitem{yu2019deep}
\bibinfo{author}{Yu, Y.}, \bibinfo{author}{Hur, T.}, \bibinfo{author}{Jung, J.} \& \bibinfo{author}{Jang, I.~G.}
\newblock \bibinfo{journal}{\bibinfo{title}{Deep learning for determining a near-optimal topological design without any iteration}}.
\newblock {\emph{\JournalTitle{Structural and Multidisciplinary Optimization}}} \textbf{\bibinfo{volume}{59}}, \bibinfo{pages}{787--799}, \doiprefix\url{https://doi.org/10.1007/s00158-018-2101-5} (\bibinfo{year}{2019}).

\bibitem{chandrasekhar2021tounn}
\bibinfo{author}{Chandrasekhar, A.} \& \bibinfo{author}{Suresh, K.}
\newblock \bibinfo{journal}{\bibinfo{title}{{TOuNN}: Topology optimization using neural networks}}.
\newblock {\emph{\JournalTitle{Structural and Multidisciplinary Optimization}}} \textbf{\bibinfo{volume}{63}}, \bibinfo{pages}{1135--1149}, \doiprefix\url{https://doi.org/10.1007/s00158-020-02748-4} (\bibinfo{year}{2021}).

\bibitem{deng2021parametric}
\bibinfo{author}{Deng, H.} \& \bibinfo{author}{To, A.~C.}
\newblock \bibinfo{journal}{\bibinfo{title}{A parametric level set method for topology optimization based on deep neural network}}.
\newblock {\emph{\JournalTitle{Journal of Mechanical Design}}} \textbf{\bibinfo{volume}{143}}, \bibinfo{pages}{091702}, \doiprefix\url{https://doi.org/10.1115/1.4050105} (\bibinfo{year}{2021}).

\bibitem{elingaard2022homogenization}
\bibinfo{author}{Elingaard, M.~O.}, \bibinfo{author}{Aage, N.}, \bibinfo{author}{B{\ae}rentzen, J.~A.} \& \bibinfo{author}{Sigmund, O.}
\newblock \bibinfo{journal}{\bibinfo{title}{De-homogenization using convolutional neural networks}}.
\newblock {\emph{\JournalTitle{Computer Methods in Applied Mechanics and Engineering}}} \textbf{\bibinfo{volume}{388}}, \bibinfo{pages}{114197}, \doiprefix\url{https://doi.org/10.1016/j.cma.2021.114197} (\bibinfo{year}{2022}).

\bibitem{yang2018mechanical}
\bibinfo{author}{Yang, L.} \emph{et~al.}
\newblock \bibinfo{journal}{\bibinfo{title}{Mechanical response of a triply periodic minimal surface cellular structures manufactured by selective laser melting}}.
\newblock {\emph{\JournalTitle{International Journal of Mechanical Sciences}}} \textbf{\bibinfo{volume}{148}}, \bibinfo{pages}{149--157}, \doiprefix\url{https://doi.org/10.1016/j.ijmecsci.2018.08.039} (\bibinfo{year}{2018}).

\bibitem{feng2022triply}
\bibinfo{author}{Feng, J.}, \bibinfo{author}{Fu, J.}, \bibinfo{author}{Yao, X.} \& \bibinfo{author}{He, Y.}
\newblock \bibinfo{journal}{\bibinfo{title}{Triply periodic minimal surface (tpms) porous structures: from multi-scale design, precise additive manufacturing to multidisciplinary applications}}.
\newblock {\emph{\JournalTitle{International Journal of Extreme Manufacturing}}} \textbf{\bibinfo{volume}{4}}, \bibinfo{pages}{022001}, \doiprefix\url{https://doi.org/10.1088/2631-7990/ac5be6} (\bibinfo{year}{2022}).

\bibitem{ding2021stl}
\bibinfo{author}{Ding, J.} \emph{et~al.}
\newblock \bibinfo{journal}{\bibinfo{title}{{STL}-free design and manufacturing paradigm for high-precision powder bed fusion}}.
\newblock {\emph{\JournalTitle{CIRP Annals}}} \textbf{\bibinfo{volume}{70}}, \bibinfo{pages}{167--170}, \doiprefix\url{https://doi.org/10.1016/j.cirp.2021.03.012} (\bibinfo{year}{2021}).

\bibitem{khaleghi2021directional}
\bibinfo{author}{Khaleghi, S.} \emph{et~al.}
\newblock \bibinfo{journal}{\bibinfo{title}{On the directional elastic modulus of the {TPMS} structures and a novel hybridization method to control anisotropy}}.
\newblock {\emph{\JournalTitle{Materials \& Design}}} \textbf{\bibinfo{volume}{210}}, \bibinfo{pages}{110074}, \doiprefix\url{https://doi.org/10.1016/j.matdes.2021.110074} (\bibinfo{year}{2021}).

\bibitem{baerentzen2005signed}
\bibinfo{author}{B{\ae}rentzen, J.~A.} \& \bibinfo{author}{Aanaes, H.}
\newblock \bibinfo{journal}{\bibinfo{title}{Signed distance computation using the angle weighted pseudonormal}}.
\newblock {\emph{\JournalTitle{IEEE Transactions on Visualization and Computer Graphics}}} \textbf{\bibinfo{volume}{11}}, \bibinfo{pages}{243--253}, \doiprefix\url{https://doi.org/10.1109/TVCG.2005.49} (\bibinfo{year}{2005}).

\bibitem{sigmund200199}
\bibinfo{author}{Sigmund, O.}
\newblock \bibinfo{journal}{\bibinfo{title}{A 99 line topology optimization code written in matlab}}.
\newblock {\emph{\JournalTitle{Structural and multidisciplinary optimization}}} \textbf{\bibinfo{volume}{21}}, \bibinfo{pages}{120--127}, \doiprefix\url{https://doi.org/10.1007/s001580050176} (\bibinfo{year}{2001}).

\bibitem{eckhart2019covalent}
\bibinfo{author}{Eckhart, K.~E.}, \bibinfo{author}{Holt, B.~D.}, \bibinfo{author}{Laurencin, M.~G.} \& \bibinfo{author}{Sydlik, S.~A.}
\newblock \bibinfo{journal}{\bibinfo{title}{Covalent conjugation of bioactive peptides to graphene oxide for biomedical applications}}.
\newblock {\emph{\JournalTitle{Biomaterials science}}} \textbf{\bibinfo{volume}{7}}, \bibinfo{pages}{3876--3885}, \doiprefix\url{https://doi.org/10.1039/C9BM00867E} (\bibinfo{year}{2019}).

\bibitem{freed2009advanced}
\bibinfo{author}{Freed, L.~E.}, \bibinfo{author}{Engelmayr~Jr, G.~C.}, \bibinfo{author}{Borenstein, J.~T.}, \bibinfo{author}{Moutos, F.~T.} \& \bibinfo{author}{Guilak, F.}
\newblock \bibinfo{journal}{\bibinfo{title}{Advanced material strategies for tissue engineering scaffolds}}.
\newblock {\emph{\JournalTitle{Advanced materials}}} \textbf{\bibinfo{volume}{21}}, \bibinfo{pages}{3410--3418}, \doiprefix\url{https://doi.org/10.1002/adma.200900303} (\bibinfo{year}{2009}).

\bibitem{lee2019resolution}
\bibinfo{author}{Lee, J.~M.}, \bibinfo{author}{Ng, W.~L.} \& \bibinfo{author}{Yeong, W.~Y.}
\newblock \bibinfo{journal}{\bibinfo{title}{Resolution and shape in bioprinting: Strategizing towards complex tissue and organ printing}}.
\newblock {\emph{\JournalTitle{Applied Physics Reviews}}} \textbf{\bibinfo{volume}{6}}, \bibinfo{pages}{011307--011322}, \doiprefix\url{https://doi.org/10.1063/1.5053909} (\bibinfo{year}{2019}).

\bibitem{dziaduszewska2021structural}
\bibinfo{author}{Dziaduszewska, M.} \& \bibinfo{author}{Zieli{\'n}ski, A.}
\newblock \bibinfo{journal}{\bibinfo{title}{Structural and material determinants influencing the behavior of porous ti and its alloys made by additive manufacturing techniques for biomedical applications}}.
\newblock {\emph{\JournalTitle{Materials}}} \textbf{\bibinfo{volume}{14}}, \bibinfo{pages}{712}, \doiprefix\url{https://doi.org/10.3390/ma14040712} (\bibinfo{year}{2021}).

\bibitem{gerhardt2010bioactive}
\bibinfo{author}{Gerhardt, L.-C.} \& \bibinfo{author}{Boccaccini, A.~R.}
\newblock \bibinfo{journal}{\bibinfo{title}{Bioactive glass and glass-ceramic scaffolds for bone tissue engineering}}.
\newblock {\emph{\JournalTitle{Materials}}} \textbf{\bibinfo{volume}{3}}, \bibinfo{pages}{3867--3910}, \doiprefix\url{https://doi.org/10.3390/ma3073867} (\bibinfo{year}{2010}).

\bibitem{chung2011design}
\bibinfo{author}{Chung, S.} \& \bibinfo{author}{King, M.~W.}
\newblock \bibinfo{journal}{\bibinfo{title}{Design concepts and strategies for tissue engineering scaffolds}}.
\newblock {\emph{\JournalTitle{Biotechnology and applied biochemistry}}} \textbf{\bibinfo{volume}{58}}, \bibinfo{pages}{423--438}, \doiprefix\url{https://doi.org/10.1002/bab.60} (\bibinfo{year}{2011}).

\bibitem{zerankeshi2022polymer}
\bibinfo{author}{Zerankeshi, M.~M.}, \bibinfo{author}{Bakhshi, R.} \& \bibinfo{author}{Alizadeh, R.}
\newblock \bibinfo{journal}{\bibinfo{title}{Polymer/metal composite 3d porous bone tissue engineering scaffolds fabricated by additive manufacturing techniques: A review}}.
\newblock {\emph{\JournalTitle{Bioprinting}}} \textbf{\bibinfo{volume}{25}}, \bibinfo{pages}{e00191}, \doiprefix\url{https://doi.org/10.1016/j.bprint.2022.e00191} (\bibinfo{year}{2022}).

\bibitem{torres2017effect}
\bibinfo{author}{Torres-Sanchez, C.} \emph{et~al.}
\newblock \bibinfo{journal}{\bibinfo{title}{The effect of pore size and porosity on mechanical properties and biological response of porous titanium scaffolds}}.
\newblock {\emph{\JournalTitle{Materials Science and Engineering: C}}} \textbf{\bibinfo{volume}{77}}, \bibinfo{pages}{219--228} (\bibinfo{year}{2017}).

\bibitem{murphy2010effect}
\bibinfo{author}{Murphy, C.~M.}, \bibinfo{author}{Haugh, M.~G.} \& \bibinfo{author}{O'brien, F.~J.}
\newblock \bibinfo{journal}{\bibinfo{title}{The effect of mean pore size on cell attachment, proliferation and migration in collagen--glycosaminoglycan scaffolds for bone tissue engineering}}.
\newblock {\emph{\JournalTitle{Biomaterials}}} \textbf{\bibinfo{volume}{31}}, \bibinfo{pages}{461--466} (\bibinfo{year}{2010}).

\bibitem{o2005effect}
\bibinfo{author}{O’Brien, F.~J.}, \bibinfo{author}{Harley, B.}, \bibinfo{author}{Yannas, I.~V.} \& \bibinfo{author}{Gibson, L.~J.}
\newblock \bibinfo{journal}{\bibinfo{title}{The effect of pore size on cell adhesion in collagen-gag scaffolds}}.
\newblock {\emph{\JournalTitle{Biomaterials}}} \textbf{\bibinfo{volume}{26}}, \bibinfo{pages}{433--441} (\bibinfo{year}{2005}).

\bibitem{SIDDIQUE2022natureInspire}
\bibinfo{author}{Siddique, S.~H.}, \bibinfo{author}{Hazell, P.~J.}, \bibinfo{author}{Wang, H.}, \bibinfo{author}{Escobedo, J.~P.} \& \bibinfo{author}{Ameri, A.~A.}
\newblock \bibinfo{journal}{\bibinfo{title}{Lessons from nature: {3D} printed bio-inspired porous structures for impact energy absorption – a review}}.
\newblock {\emph{\JournalTitle{Additive Manufacturing}}} \textbf{\bibinfo{volume}{58}}, \bibinfo{pages}{103051}, \doiprefix\url{https://doi.org/10.1016/j.addma.2022.103051} (\bibinfo{year}{2022}).

\bibitem{BANDYOPADHYAY2021NatureInspire}
\bibinfo{author}{Bandyopadhyay, A.}, \bibinfo{author}{Traxel, K.~D.} \& \bibinfo{author}{Bose, S.}
\newblock \bibinfo{journal}{\bibinfo{title}{Nature-inspired materials and structures using {3D} printing}}.
\newblock {\emph{\JournalTitle{Materials Science and Engineering: R: Reports}}} \textbf{\bibinfo{volume}{145}}, \bibinfo{pages}{100609}, \doiprefix\url{https://doi.org/10.1016/j.mser.2021.100609} (\bibinfo{year}{2021}).

\bibitem{al2019multifunctional}
\bibinfo{author}{Al-Ketan, O.} \& \bibinfo{author}{Abu Al-Rub, R.~K.}
\newblock \bibinfo{journal}{\bibinfo{title}{Multifunctional mechanical metamaterials based on triply periodic minimal surface lattices}}.
\newblock {\emph{\JournalTitle{Advanced Engineering Materials}}} \textbf{\bibinfo{volume}{21}}, \bibinfo{pages}{1900524}, \doiprefix\url{https://doi.org/10.1002/adem.201900524} (\bibinfo{year}{2019}).

\bibitem{xu2024TPMS}
\bibinfo{author}{Xu, W.} \emph{et~al.}
\newblock \bibinfo{journal}{\bibinfo{title}{Topology optimization via spatially-varying {TPMS}}}.
\newblock {\emph{\JournalTitle{IEEE Trans. on Visualization and Computer Graphics}}} \textbf{\bibinfo{volume}{30}}, \bibinfo{pages}{4570--4587}, \doiprefix\url{https://doi.org/10.1109/TVCG.2023.3268068} (\bibinfo{year}{2024}).

\bibitem{liu2014efficient}
\bibinfo{author}{Liu, K.} \& \bibinfo{author}{Tovar, A.}
\newblock \bibinfo{journal}{\bibinfo{title}{An efficient {3D} topology optimization code written in {Matlab}}}.
\newblock {\emph{\JournalTitle{Structural and multidisciplinary optimization}}} \textbf{\bibinfo{volume}{50}}, \bibinfo{pages}{1175--1196}, \doiprefix\url{https://doi.org/10.1007/s00158-014-1107-x} (\bibinfo{year}{2014}).

\bibitem{SigmundEtAl1998NumericalInstabilities}
\bibinfo{author}{Sigmund, O.} \& \bibinfo{author}{Petersson, J.}
\newblock \bibinfo{journal}{\bibinfo{title}{Numerical instabilities in topology optimization: a survey on procedures dealing with checkerboards, mesh-dependencies and local minima}}.
\newblock {\emph{\JournalTitle{Structural optimization}}} \textbf{\bibinfo{volume}{16}}, \bibinfo{pages}{68--75}, \doiprefix\url{https://doi.org/10.1007/BF01214002} (\bibinfo{year}{1998}).

\bibitem{bendsoe1995optimization}
\bibinfo{author}{Bends{\o}e, M.~P.}
\newblock \emph{\bibinfo{title}{Optimization of structural topology, shape, and material}}, vol. \bibinfo{volume}{414} (\bibinfo{publisher}{Springer}, \bibinfo{year}{1995}).

\bibitem{svanberg1987method}
\bibinfo{author}{Svanberg, K.}
\newblock \bibinfo{journal}{\bibinfo{title}{The method of moving asymptotes—a new method for structural optimization}}.
\newblock {\emph{\JournalTitle{International journal for numerical methods in engineering}}} \textbf{\bibinfo{volume}{24}}, \bibinfo{pages}{359--373}, \doiprefix\url{https://doi.org/10.1002/nme.1620240207} (\bibinfo{year}{1987}).

\bibitem{kingma2014adam}
\bibinfo{author}{Kingma, D.~P.} \& \bibinfo{author}{Ba, J.}
\newblock \bibinfo{journal}{\bibinfo{title}{Adam: A method for stochastic optimization}}.
\newblock {\emph{\JournalTitle{arXiv preprint arXiv:1412.6980}}} \doiprefix\url{https://doi.org/10.48550/arXiv.1412.6980} (\bibinfo{year}{2014}).

\bibitem{fiedler2015mechanical}
\bibinfo{author}{Fiedler, T.} \emph{et~al.}
\newblock \bibinfo{journal}{\bibinfo{title}{On the mechanical properties of plc--bioactive glass scaffolds fabricated via bioextrusion}}.
\newblock {\emph{\JournalTitle{Materials Science and Engineering: C}}} \textbf{\bibinfo{volume}{57}}, \bibinfo{pages}{288--293}, \doiprefix\url{https://doi.org/10.1016/j.msec.2015.07.063} (\bibinfo{year}{2015}).

\end{thebibliography}
